# All-optical room-temperature spin coherence of aqueous asymmetrically chelated Cu(II) for chemoselective discrimination

Jinlei Li[1], Qiaoyun Wu[1], Qiaoqiao Shao[1], Zhubin Hu[1], Haitao Sun[1,2], Zhenrong Sun[1,3,*], Donghai Feng[1,3,*]

[1]*State Key Laboratory of Precision Spectroscopy, School of Physics, East China Normal University, Shanghai 200241, China*

[2]*AI Physics Interdisciplinary Research Center (AIPIRC), School of Physics,*

*East China Normal University, Shanghai 200241, China*

[3]*Collaborative Innovation Center of Extreme Optics, Shanxi University, Taiyuan, Shanxi 030006, China*

Email: zrsun@phy.ecnu.edu.cn; dhfeng@phy.ecnu.edu.cn

**Abstract:**

Harnessing molecular electron spins as responsive quantum sensors in ambient water requires bridging ultrafast all-optical readout with selective chemical recognition. Here, we overcome the intrinsic magneto-optical sluggishness of 3d transition-metal complexes—which previously restricted all-optical detection to high concentrations of ~1 M—through a coordination-driven symmetry-breaking strategy. Coordinating aqueous Cu(II) with EDTA imposes a non-centrosymmetric ligand field with enhanced metal–ligand covalency, relaxing Laporte restrictions and amplifying d–d transition dipole moments. This design enables room-temperature all-optical initialization and readout of electron spin coherence down to 100 μM with an apparent dephasing time $T_2^* \approx 0.54$ ns. Crucially, the chelated complex chemoselectively discriminates amino acid biomolecules: histidine binding induces a pronounced, concentration-dependent prolongation of $T_2^*$ to ~0.8 ns, whereas glycine produces no measurable effect. These results transform spin dephasing into a chemically informative observable, opening avenues toward microwave-free, analyte-responsive quantum sensing in complex aqueous environments.

## Introduction

Molecular electron spins offer a distinct paradigm at the interface of quantum coherence and synthetic chemistry. A diverse family of molecular spins—including stable organic radicals[1–3], photoexcited triplet states[4,5], and metal coordination complexes[6–8]—readily supports electron spin within structures that are synthetically tunable, solution-processable, and chemically addressable. Among these classes, transition-metal coordination complexes are particularly well suited to establish a chemoselective quantum sensing framework. Because coordination chemistry—rather than a rigid host lattice—defines the spin-bearing site, the spin Hamiltonian can be both synthetically pre-programmed and dynamically modulated by coordination events. This provides a route to selectively transduce chemical binding events into quantum metrics, exploiting the intimate dependence of spin coherence on the local coordination environment as an active readout.

By contrast, nitrogen–vacancy (NV) centres in diamond—a leading platform for nanoscale quantum sensing[9,10]—do not possess an intrinsic coordination site for molecular recognition. Although nanoscale NMR can in principle resolve nuclear-spin signatures of diamagnetic biomolecules[11], practical chemical discrimination typically requires surface functionalization, immobilized synthetic receptors, or extrinsic spin labels[12,13]. NV centres thus detect analytes primarily via stray electric and magnetic perturbations, rather than through direct modulations of an intrinsic ligand field. Consequently, chemical selectivity cannot be derived from the diamond lattice itself and must instead be engineered externally at the diamond–analyte interface.

Exploiting spin dynamics to track coordination events in ambient solutions, however, requires detection techniques capable of resolving ultrafast processes. Pulsed electron paramagnetic resonance is inherently constrained by cavity bandwidth and, critically, by instrumental dead times that obscure the picosecond-to-nanosecond dephasing of molecular spins in solution[14]. Ultrafast all-optical pump–probe spectroscopy circumvents these limits, enabling spin initialization and readout on sub-picosecond timescales. Recent demonstrations of all-optical coherence in hexahaloiridate complexes have validated this approach and extended it into the tissue-transparency window[15,16]. Yet these systems possess coordinatively saturated coordination spheres that are chemically inert under physiological conditions: they excel at reporting macroscopic variables such as viscosity, but lack the labile coordination sites required for specific molecular recognition. Moreover, the Ir(IV) oxidation state introduces potential redox incompatibility with redox-sensitive biomolecules. Consequently, despite these pioneering advances, an all-optical molecular spin probe that directly couples chemical recognition to spin coherence has remained elusive.

Here we address this challenge with Cu(II), a canonical $S$ = 1/2 transition-metal centre whose $d^9$ configuration provides a chemically labile, biomolecule-receptive coordination environment. Unlike rigid solid-state defects or chemically inert hexahaloiridates, Cu(II) ions coordinate readily to amino acids, peptides, and proteins, offering a direct chemical handle for biological recognition[17]. However, all-optical spin detection of Cu(II) in aqueous solution is severely limited by weak magneto-optical signals. The orbitally non-degenerate ground state leads to quenching of orbital angular momentum, and the weak d–d transition strength of the fluxional aqua shell strongly limits transition dipole moments, attenuating the efficiency of both spin initialization and readout. Previous all-optical measurements on aqueous Cu(II) aqua ions were conducted at concentrations in the 0.7–1.4 M range[18]. To overcome the intrinsically weak transition dipole moments of fluxional aqua complexes, we introduce a symmetry-breaking chelation strategy: coordinating Cu(II) with EDTA imposes structural asymmetry and enhances metal–ligand covalency. This relaxes the Laporte selection rule, yielding a 7.3-fold enhancement in d–d absorption and a 5.8-fold amplification in magnetic circular dichroism (MCD). Benefiting from these augmented transition dipole moments, high-sensitivity polarization-resolved spectroscopy enables all-optical spin readout down to 100 μM (and extending to 20 μM at boosted probe power) in room-temperature water.

Crucially, the EDTA-chelated Cu(II) complex retains a chemically responsive coordination sphere. Histidine binding modulates local symmetry and solvation, shifting the dephasing time $T_2^*$ from 0.54 ns to ~0.8 ns in a concentration-dependent manner, whereas glycine produces no detectable shift. The coherence modulation therefore reports specific coordination rather than non-specific solvent effects. These results recast spin dephasing—usually treated as environmental noise—as a chemically informative observable. By converting coordination dynamics into a spectroscopic signature of analyte identity, we translate the broad concept of harnessing decoherence into a viable framework for chemoselective molecular recognition.

## Results and discussion

### Symmetry-breaking coordination amplifies optical transitions

A central challenge in optical spin manipulation of 3d transition-metal ions is the weak oscillator strength of d–d transitions under centrosymmetric or near-centrosymmetric coordination. Although the hydrated Cu(II) ion has traditionally been described as a Jahn–Teller-elongated octahedral $[Cu(H_2O)_6]^{2+}$ complex, neutron diffraction and *ab initio* molecular dynamics demonstrate that its first hydration shell is predominantly five-coordinate rather than a static octahedron[19]. X-ray absorption spectroscopy further reveals its local geometry as an axially elongated square pyramid, $[Cu(H_2O)_5]^{2+}$, with four short equatorial and one long axial Cu–O bonds[20,21]. Despite picosecond dynamical interconversion to trigonal bipyramidal configurations, this square-pyramidal motif represents the time-averaged, energetically dominant species in aqueous solution and was therefore adopted as the representative model for our quantum chemical calculations (Fig. 1a). Even in this five-coordinate geometry, the homoleptic aqua ligands provide a weak, pseudo-centrosymmetric ligand field with minimal metal–ligand covalency; consequently, the d–d transitions remain largely Laporte-forbidden—deriving only vanishingly small intensity from weak vibronic borrowing and residual static symmetry breaking—resulting in negligible absorption across the visible and near-infrared regions (Supplementary Notes).

Upon chelation with EDTA, the Cu(II) centre is coordinated by amine nitrogen and carboxylate oxygen donors in a rigid, non-centrosymmetric framework, predominantly adopting a hexadentate form in dynamic equilibrium with an aquo-pentadentate species in aqueous solution (Fig. 1b and Supplementary Notes)[22]. This pronounced structural asymmetry, coupled with enhanced metal–ligand orbital mixing—as supported by increased ligand participation in the hole–electron distribution and higher coordination bond orders (Supplementary Fig. 1)—relaxes the Laporte restriction and markedly increases the electric-dipole intensity of the predominantly ligand-field transitions. Concurrently, the stronger ligand field enlarges the d-orbital splitting, shifting the absorption profile to higher energies. Experimentally, aqueous $Na_2[Cu(EDTA)]$ (Cu–EDTA) displays a prominent d–d band centred at ~735 nm (Fig. 1c, red), delivering a 7.3-fold enhancement in molar absorptivity relative to aqueous $CuSO_4$ (Fig. 1c, blue). This symmetry-breaking effect is mirrored in MCD spectroscopy: the chelated complex exhibits a strong differential molar absorptivity (Fig. 1d, red), representing a 5.8-fold amplification in peak MCD amplitude over the unchelated aqua ion (Fig. 1d, blue). First-principles quantum chemical calculations (TD-DFT and multireference QD-SC-NEVPT2) capture the overall magnitude and trend of these optical and magneto-optical enhancements when the coordination equilibrium is taken into account (Supplementary Notes, Supplementary Fig. 2 and Supplementary Table 1). These augmented optical responses validate our coordination-engineering strategy, establishing an effective spectroscopic pathway for all-optical initialization and readout of Cu(II) spin states with elevated sensitivity at room temperature.

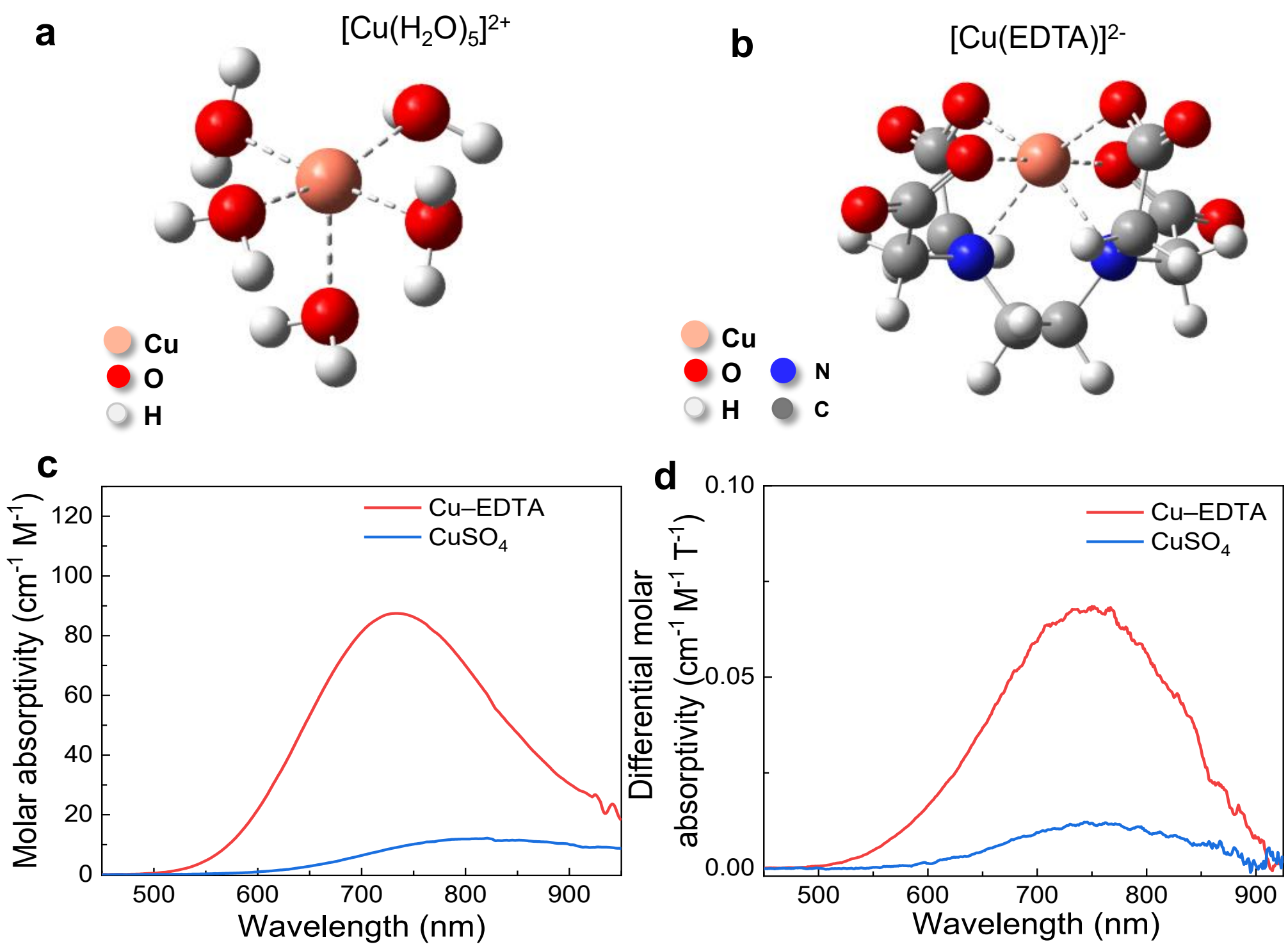

**Fig. 1 | Coordination structures and optical spectra of aqueous $Cu^{2+}$ complexes.** **a**, DFT-optimized geometry of the representative aqua copper ion ($[Cu(H_2O)_5]^{2+}$, square pyramidal). **b**, DFT-optimized geometry of the six-coordinate $[Cu(EDTA)]^{2-}$ complex. **c**, UV–vis absorption spectra of aqueous $Na_2[Cu(EDTA)]$ (Cu–EDTA) and $CuSO_4$ solutions. **d**, Corresponding MCD spectra. Computational details and simulated spectra are provided in Supplementary Notes and Supplementary Fig. 2.

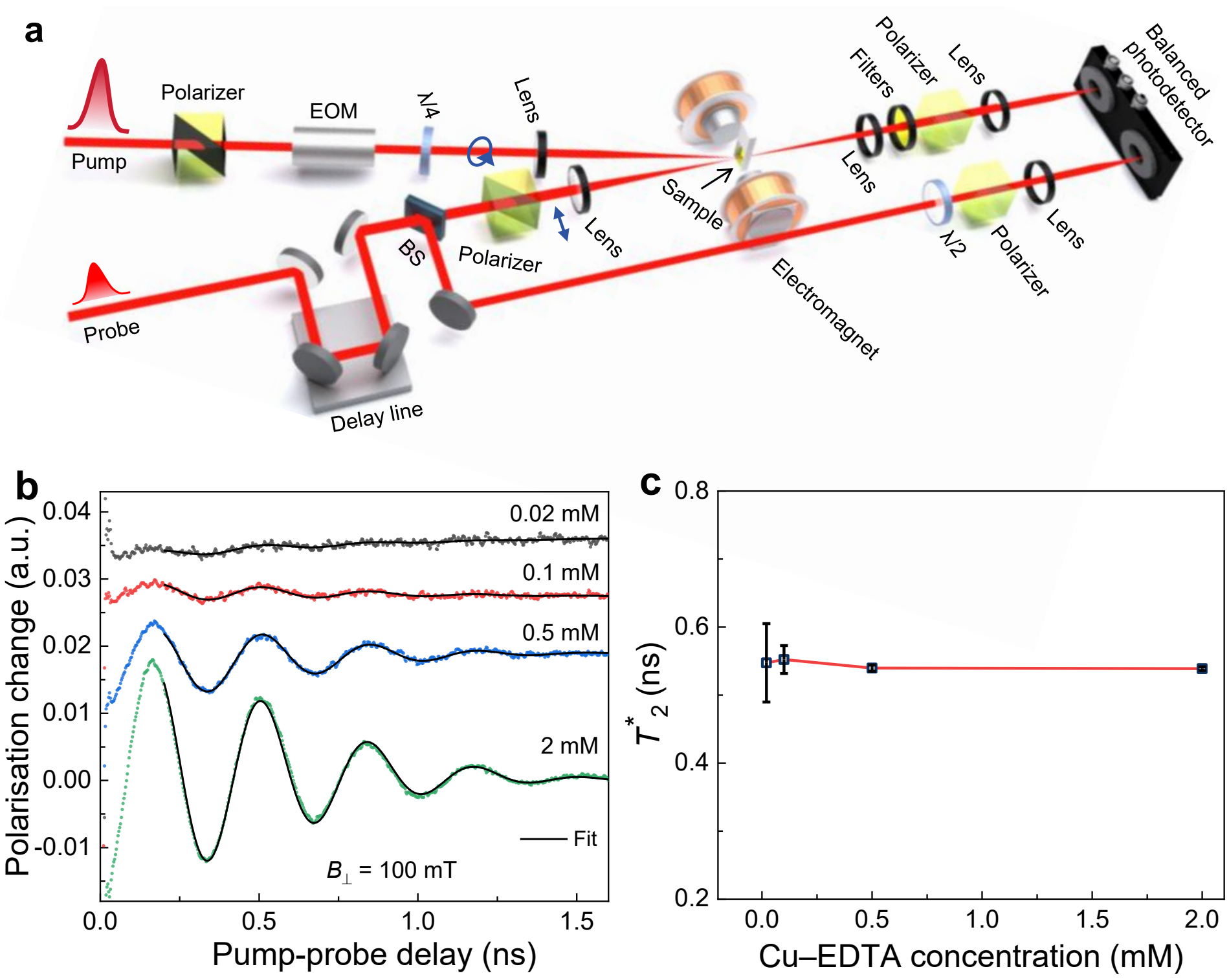

**Fig. 2 | Ultrafast optical detection and concentration-dependent spin coherence dynamics in aqueous Cu–EDTA. a**, Schematic of the non-degenerate polarisation-sensitive pump–probe setup. Spin coherence is generated by a circularly polarised, 750 nm pump pulse modulated between $\sigma^+$ and $\sigma^-$ via an electro-optic modulator (EOM), and monitored through transient polarisation changes of a time-delayed, 690 nm probe pulse using crossed polarisers and balanced lock-in detection. Residual 750 nm pump light is blocked by spectral filters before detection. BS, beam splitter; $\lambda/4$, quarter-wave plate; $\lambda/2$, half-wave plate. **b**, Time-resolved spin coherence dynamics in aqueous Cu–EDTA solutions across concentrations from 0.02 to 2 mM recorded under a transverse magnetic field of $B_\perp = 100$ mT. Traces are vertically offset for clarity. Probe power was set to 20 mW (0.1–2 mM) and 100 mW (0.02 mM) to optimize the signal-to-noise ratio. Solid black curves represent fits to $S(t) \propto \cos(2\pi\nu t)\exp\left[-t^2/\left(2T_2^{*2}\right)\right]$ within the fitting window $t \geq 0.2$ ns to exclude the instantaneous optical Kerr response near zero delay (Supplementary Fig. 3). **c**, Spin dephasing time ($T_2^*$) as a function of Cu–EDTA concentration. The red line is a guide to the eye; error bars represent the fitting standard errors.

**All-optical spin readout down to micromolar concentrations**

To explore the electron spin dynamics of aqueous Cu–EDTA, we implemented polarisation-sensitive pump–probe spectroscopy in a non-degenerate geometry (Fig. 2a). A circularly polarised 750 nm pump pulse initialises spin polarisation, while a linearly polarised 690 nm probe monitors the spin-induced optical polarisation rotation and ellipticity. Detection using crossed Glan prisms and a balanced receiver isolated the spin signals while suppressing the background transmission, enabling the use of high probe powers to detect weak signals without detector saturation (Methods).

The low-symmetry Cu–EDTA chromophore provides a viable optical readout handle at room temperature. Under a modest transverse magnetic field of $100\ \mathrm{mT}$, distinct coherent spin precession oscillations persist into the nanosecond regime across two orders of magnitude in concentration ($0.02 - 2\ \mathrm{mM}$, Fig. 2b). The oscillating spin coherence is well captured by a Gaussian-damped cosine function:

$$S(t) \propto \cos(2\pi\nu t)\exp\left(-\frac{t^2}{2T_2^{*2}}\right) \quad (1)$$

where $\nu$ is the Larmor frequency and $T_2^*$ is the spin dephasing time. To exclude non-resonant solvent response and optical Kerr artifacts around time zero (Supplementary Fig. 3), the fits were performed over the evaluation window $t \geq 0.2\ \mathrm{ns}$ while keeping the pump–probe time origin ($t = 0$) fixed. By contrast, a mono-exponential damping profile fails to match the experimental envelope (Supplementary Fig. 4).

Strikingly, the coherent spin response remains clearly discernible down to $100\ \mu\mathrm{M}$ under standard probe conditions, and extends to $0.02\ \mathrm{mM}$ ($20\ \mu\mathrm{M}$) upon increasing probe power (Fig. 2b)—a sensitivity floor inaccessible to conventional optical spin probing of aqueous Cu(II) and comparable to micromolar benchmarks recently reported for the heavy transition-metal system $[\mathrm{IrCl_6}]^{2-}$ using all-optical detection[15].

Across this broad concentration span ($0.02 - 2$ mM), the extracted $T_2^*$ remains essentially invariant at $\approx 0.54$ ns (Fig. 2c). This strict concentration invariance rules out intermolecular Cu–Cu dipolar coupling (or dipolar-driven spin cross relaxation[18]) and concentration-dependent aggregation as dominant dephasing channels.

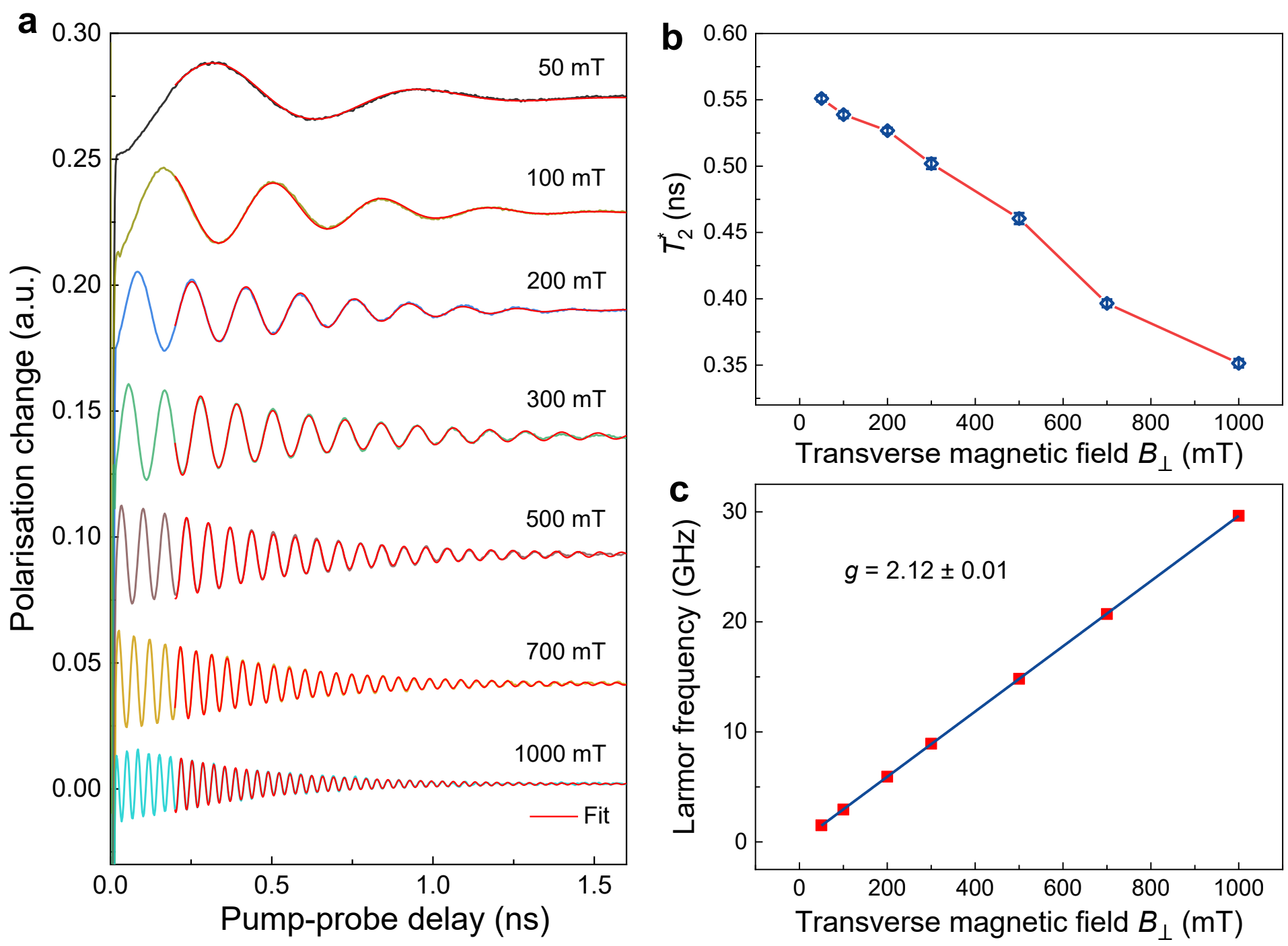

**Fig. 3 | Magnetic-field dependence of all-optical spin coherence in aqueous Cu–EDTA.** **a**, Time-resolved polarization change traces measured for a 2 mM aqueous solution of Cu–EDTA under external transverse magnetic fields ($B_\perp$) from 50 to 1000 mT. Traces are vertically offset for clarity. To avoid the optical Kerr response near zero delay, fits were performed over $t \geq 0.2$ ns using a Gaussian-damped cosine function ($S(t) \propto \cos(2\pi\nu t)\exp\left[-t^2/\left(2T_2^{*2}\right)\right]$) for 50–200 mT and a mono-exponentially damped cosine function ($S(t) \propto \cos(2\pi\nu t)\exp(-t/T_2^*)$) for 300–1000 mT (solid red curves). **b**, Transverse spin dephasing time ($T_2^*$) as a function of the external magnetic field, extracted from the fits in **a**. The solid red line is a guide to the eye; error bars represent the fitting standard errors. **c**, Larmor precession frequency plotted against the magnetic field. The solid blue line is a linear fit constrained through the origin ($\nu = g\mu_B B_\perp/h$), yielding an effective isotropic Landé $g$-factor of $2.12 \pm 0.01$.

## Transverse-field-dependent spin dephasing dynamics

Having established all-optical readability down to the micromolar regime and concentration-stable coherence in the EDTA-chelated Cu(II) system, we next employed the transverse magnetic field ($B_{\perp}$) as a quantitative control parameter to elucidate the magnetic origin of the spin response. At a fixed benchmark concentration of 2 mM, the transient ellipticity traces evolve systematically with increasing $B_{\perp}$ (Fig. 3a): low fields produce slowly oscillating spin precession signals, whereas fields up to 1000 mT induce rapid oscillations with pronounced damping.

Consistent with the behavior detailed in Fig. 3a, a magnetic-field-dependent fitting analysis reveals a smooth phenomenological crossover: while a Gaussian damping function best captures the dephasing oscillations at lower fields (50–200 mT; Supplementary Fig. 4) and a mono-exponential decay, $S(t) \propto \cos(2\pi\nu t)\exp(-t/T_2^*)$, provides superior fidelity at higher fields (700–1000 mT; Supplementary Fig. 5), the intermediate regime (300–500 mT) represents a gradual crossover where both models yield comparable residuals (Supplementary Fig. 6). Traces are modeled with Gaussian (50–200 mT) and mono-exponential (300–1000 mT) envelopes, yielding a spin dephasing time that decreases monotonically from $\approx 0.55$ ns at 50 mT to $\approx 0.35$ ns at 1000 mT (Fig. 3b). The low-field Gaussian envelope is a hallmark of dephasing driven by nuclear hyperfine baths in localized spin systems[23–25], here associated with local hyperfine fields of the inner coordination sphere (substantiated in Fig. 5). At elevated fields, this regime gives way to accelerated dephasing driven by Zeeman-amplified inhomogeneous broadening ( $1/T_2^* \propto \Delta g \cdot \mu_B B_{\perp}/\hbar$ ) arising from the distribution of effective $g$-factors (governed by orientational dispersion and local $g$- strain) in the solution ensemble. A linear fit of the dephasing rate against the transverse magnetic field yields an effective $g$-factor spread of $\Delta g = 0.012$ (Supplementary Fig. 7).

The precession frequency scales strictly linearly with the magnetic field (Fig. 3c). A linear fit through the origin ($\nu = g_{\mathrm{iso}}\mu_{\mathrm{B}}B_{\perp}/h$) yields an effective isotropic Landé factor of $g_{\mathrm{iso}} = 2.12 \pm 0.01$. This value is slightly lower than that experimentally measured for aqueous $CuSO_4$ ($g_{\mathrm{iso}} = 2.20 \pm 0.05$[18]), fully consistent with the enhanced ligand-field splitting and increased covalency imparted by EDTA chelation in a dynamically averaged, Jahn–Teller-distorted coordination environment, confirming that the transient polarization oscillations originate from the coherent spin precession of Cu(II) electron spins.

**Chemoselective coherence modulation by histidine**

With the magnetic-field dependence of Cu–EDTA established, we explored the susceptibility of the optical spin coherence to local chemical and biomolecular perturbations. At a constant field of $B_{\perp} = 50$ mT, the addition of L-histidine (His) to a 2 mM Cu–EDTA aqueous solution induces a pronounced, concentration-dependent deceleration of spin dephasing in the time-domain precession traces (Fig. 4a and Supplementary Fig. 8). Quantitative parameterization reveals that the transverse dephasing time, $T_2^*$, increases linearly with histidine concentration in the low-concentration regime (0–12 mM) as $T_2^* = 0.017c_{\mathrm{His}} + 0.53$ ns (Fig. 4b). This pronounced response ($\sim 17$ ps mM$^{-1}$) elevates $T_2^*$ to $\sim 0.76$ ns, beyond which the response markedly levels off into a quasi-saturation regime spanning ~0.78–0.83 ns at 40–100 mM (Fig. 4b).

Because EDTA firmly saturates the primary inner coordination sphere, this prolongation is governed by outer-sphere supramolecular interactions—primarily hydrogen bonding and transient association involving the imidazole moiety within the second coordination sphere. This outer-sphere mechanism is directly corroborated by steady-state absorption spectroscopy (Supplementary Fig. 9): across the concentration range of $T_2^*$ prolongation ($\leq 20$ mM His), the characteristic d–d absorption band of Cu–EDTA exhibits no peak shift ($\Delta\lambda_{\max} = 0$) and only a minor reduction in absorption intensity, ruling out direct inner-sphere ligand substitution. The detailed microscopic mechanism governing the $T_2^*$ prolongation will be addressed in the subsequent longitudinal-field dependence analysis.

By contrast, control measurements utilizing glycine (Gly)—which lacks the aromatic imidazole moiety—produce a virtually flat response (Supplementary Fig. 10). Time-resolved precession traces recorded in the presence of 10 mM Gly overlay closely with pristine Cu–EDTA (Fig. 4c). Across 0–10 mM Gly, $T_2^*$ remains essentially pinned within 0.53–0.54 ns without a discernible upward trajectory (Fig. 4d), matching steady-state spectra where Gly leaves both the absorption maximum and intensity unchanged (Supplementary Fig. 9). This direct side-by-side comparison confirms that the all-optical coherence readout does not merely register generic ionic strength or solution viscosity variations, but discriminates subtle differences in amino-acid coordination chemistry. We emphasize that these findings constitute a proof-of-principle for molecular spin-based chemical discrimination: the divergent dephasing trajectories between histidine and glycine demonstrate the chemical sensitivity accessible to this all-optical Cu(II) spin probe, while full analytical benchmarking and sensing optimization remain subjects for future investigation.

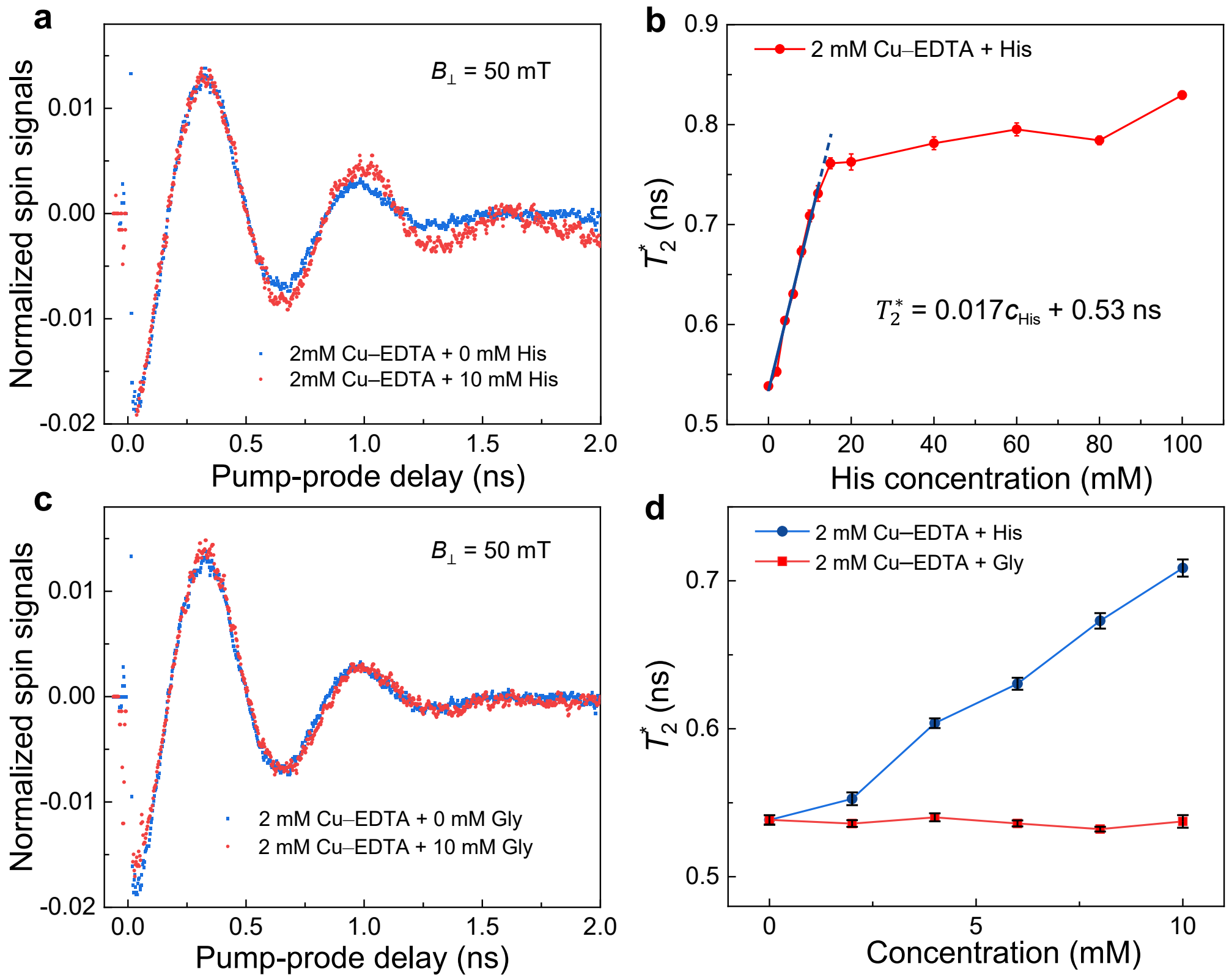

**Fig. 4 | Specific prolongation of Cu–EDTA spin coherence induced by histidine. a**, Time-resolved spin precession traces for 2 mM Cu–EDTA in the absence (blue) and presence (red) of 10 mM histidine (His), showing noticeable slowing of spin dephasing upon His addition. **b**, Transverse spin dephasing time ($T_2^*$) as a function of His concentration (0–100 mM). Dashed lines represent linear fits illustrating the initial concentration-dependent prolongation and the subsequent quasi-saturation; the red solid curve is a guide to the eye. **c**, Time-resolved spin precession traces for 2 mM Cu–EDTA without glycine (Gly, blue) and with 10 mM Gly (red), showing negligible influence of Gly on the dephasing dynamics. **d**, Comparison of $T_2^*$ evolution as a function of amino acid concentration between His (blue circles) and Gly (red squares) up to 10 mM in 2 mM Cu–EDTA solutions (solid lines are guides to the eye). In **a** and **c**, scatter points represent measured raw spin signals. In **b** and **d**, data points and error bars correspond to the best-fit values and standard errors extracted from Gaussian-damped cosine fits of individual precession traces (Supplementary Fig. 8 and Fig. 10).

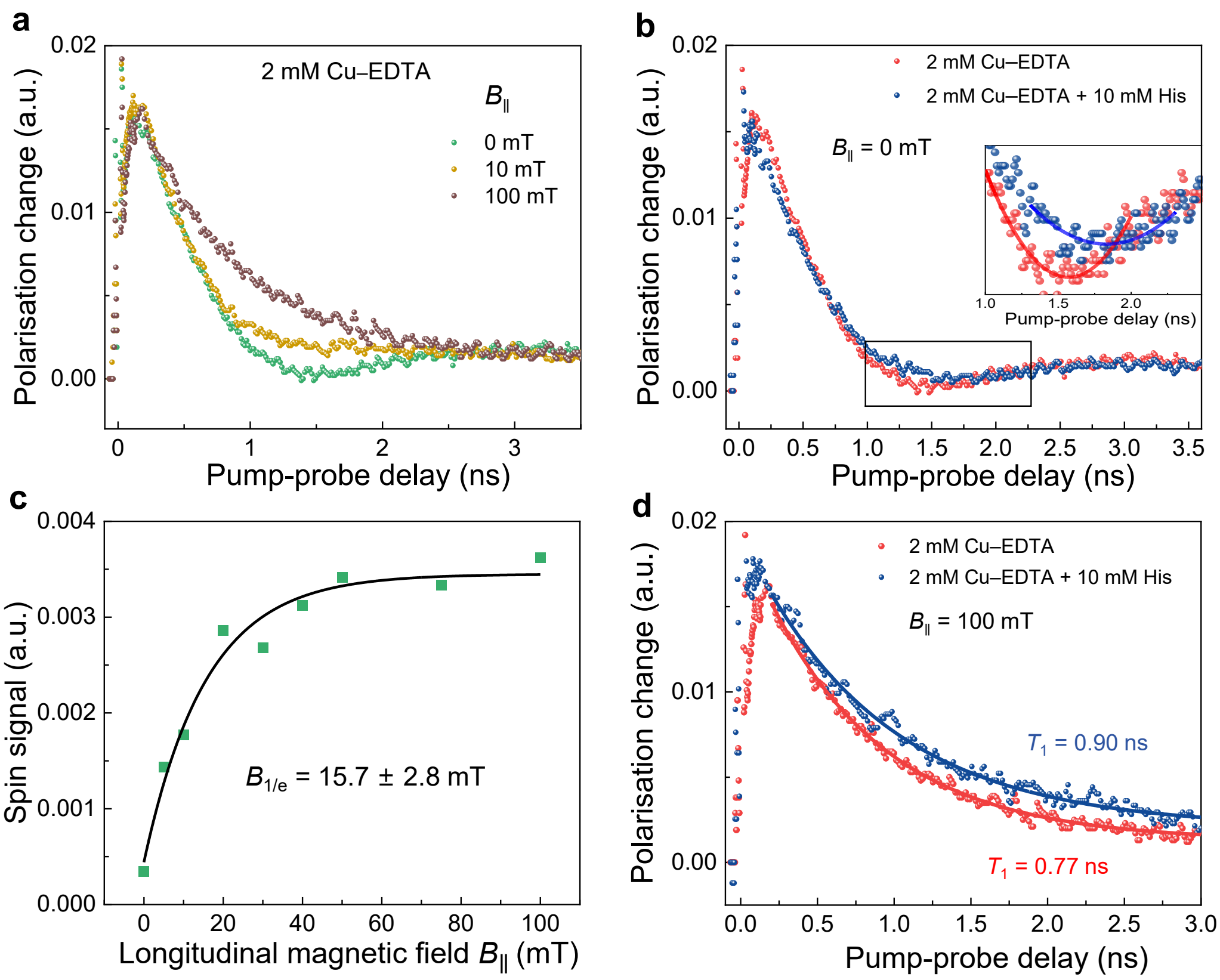

**Fig. 5 | Magnetic-field-dependent spin dynamics and histidine coordination of aqueous Cu–EDTA. a**, Spin dynamics of 2 mM Cu–EDTA under representative longitudinal magnetic fields ($B_{\parallel} = 0,\ 10$, and $100$ mT; see Extended Data Fig. 1 for the full magnetic-field series). At zero field, a characteristic dip appears near 1.57 ns followed by a partial recovery. **b**, Zero-field spin dynamics of 2 mM Cu–EDTA in the absence and presence of 10 mM histidine (His). Inset: enlarged view of the boxed region showing the rightward shift of the minimum upon His coordination. Solid curves are phenomenological cubic polynomial fits across the trough regions (1.0–2.0 ns for Cu–EDTA; 1.3–2.3 ns for Cu–EDTA + His) used to determine the minimum positions ($t_{\mathrm{min}} = 1.57 \pm 0.01$ ns and $1.83 \pm 0.02$ ns, respectively). **c**, Average spin signal across the zero-field trough region (1.50–1.64 ns) as a function of longitudinal magnetic field ($B_{\parallel}$) extracted from **a** and Extended Data Fig. 1. The solid curve is an exponential saturation fit, yielding a characteristic field $B_{1/\mathrm{e}} = 15.7 \pm 2.8$ mT. **d**, Spin dynamics of 2 mM Cu–EDTA with and without 10 mM His at $B_{\parallel} = 100$ mT. Solid curves are single-exponential fits, yielding the indicated spin lifetimes.

## Longitudinal-field spin dynamics and nuclear bath decoupling

To uncover the microscopic mechanisms governing spin coherence in Cu–EDTA and its modulation by outer-sphere coordination, we next investigated the spin dynamics under zero and longitudinal magnetic fields. The zero-field spin dynamics of aqueous Cu–EDTA exhibit a pronounced non-monotonic profile: a rapid initial collapse toward a minimum at $\sim 1.57$ ns, followed by a discernible recovery into an extended, decaying plateau (green trace in Fig. 5a and red trace in Fig. 5b). While such transient signatures are quintessential features of localized spins in solid-state matrices—such as trivalent rare-earth ions[23] and colloidal semiconductor quantum dots[24]—their clear observation in fluid solution is extraordinary. Phenomenologically, the observed trajectory conforms closely to the Merkulov–Efros–Rosen (MER) formalism originally developed for localized electronic spins coupled to an ensemble of nuclear spins[25,26]:

$$\langle S(t)\rangle = \frac{S_0}{3}\left[1 + 2\left(1 - 2\left(\frac{t}{2T_\Delta}\right)^2\right)\exp\left(-\left(\frac{t}{2T_\Delta}\right)^2\right)\right], \quad (2)$$

where $S_0$ is the initial spin signal, and $T_\Delta = \hbar/(g\mu_B\Delta_B)$ represents the characteristic dephasing timescale governed by the root-mean-square amplitude of the local nuclear hyperfine fluctuation field, $\Delta_B$. Although this continuum formalism was originally derived for static nuclear spin ensembles in the solid state, it serves here as a phenomenological model and an effective coarse-grained metric to parameterize the local hyperfine interaction. Under this framework, where the polarization minimum occurs at $t_{\min} = \sqrt{6}T_\Delta$, the observed trough at $\sim 1.57$ ns yields an apparent dephasing timescale $T_\Delta = t_{\min}/\sqrt{6} \approx 0.64$ ns, which corresponds to an effective local hyperfine field of $\Delta_B = \hbar/(g\mu_B T_\Delta) \approx 8.4$ mT (using $g = 2.12$).

The nuclear environment has been extensively modeled as a primary source of electron spin decoherence in transition metal complexes[27,28]. In the absence of an overriding Zeeman field, the unquenched hyperfine coupling of the central metal ($^{63,65}$Cu) and coordinated donor nuclei ($^{14}$N) acts as an effective quasi-static local field, triggering rapid hyperfine dephasing—a bottleneck that can typically be bypassed by exhaustive ligand/matrix nuclear spin engineering or isotopic purification[29,30]. Specifically, dynamical decoupling noise spectroscopy on molecular copper qubits directly resolves the discrete environmental noise bath generated by the $^{63}$Cu nuclear spin alongside surrounding ligands[31]. In Cu(II) complexes, the intrinsic copper hyperfine splitting typically spans tens to hundreds of megahertz ($A_{\parallel} \sim 400-600$ MHz and $A_{\perp} \sim 50-150$ MHz, corresponding to ~15–20 mT and ~2–5 mT, respectively)[31,32]. In ambient aqueous solution, fast local conformational fluctuations partially average out the anisotropic dipolar components (operating on picosecond timescales); however, the isotropic Fermi contact coupling survives as an effective quasi-static random field with a net root-mean-square amplitude $\Delta_B$ of a few milliteslas (consistent with $\Delta_B \approx 8.4$ mT derived above), quantitatively matching the emergence of the dephasing minimum at 1.57 ns (green trace in Fig. 5a and red trace in Fig. 5b).

Intriguingly, comparative measurements performed in $H_2O$ and $D_2O$ reveal virtually identical spin dephasing dynamics (Supplementary Fig. 11). Considering that the gyromagnetic ratio of deuterium is nearly an order of magnitude smaller than that of protium ($\gamma_D/\gamma_H \approx 0.15$), the absence of a solvent isotope effect unambiguously indicates that outer-sphere solvent diffusion contributes negligibly to the nuclear dephasing bath. Instead, low-field dephasing is governed by the immediate inner coordination sphere consisting of the central $^{63,65}$Cu and chelating core, demonstrating that spin dephasing at low fields is dictated by the primary coordination framework and its local nuclear hyperfine fields rather than the nuclear spin bath of the bulk solvent.

Applying an external longitudinal magnetic field ($B_{\parallel}$) progressively decouples the electron spin from this inner-sphere nuclear bath (Fig. 5a). With increasing $B_{\parallel}$, the characteristic dephasing dip is systematically suppressed, and the residual polarization within the trough window (1.50–1.64 ns) recovers monotonically (Fig. 5c). This field-dependent recovery is well parameterized by an exponential saturation profile, yielding a characteristic field of $B_{1/\mathrm{e}} = 15.7 \pm 2.8$ mT (Fig. 5c). Notably, while the effective local fluctuation field $\Delta_B \approx 8.4$ mT dictates the zero-field dephasing timescale, complete decoupling requires overriding the dominant anisotropic coupling of the inner-sphere $^{63,65}$Cu core ($A_{\parallel} \sim 15 - 20$ mT). The extracted $B_{1/\mathrm{e}}$ quantitatively coincides with this $A_{\parallel}$ threshold, confirming that the recovery is governed by the systematic quenching of inner-sphere hyperfine channels. This crossover behavior conceptually parallels the magnetic field effects observed in molecular copper charge-transfer architectures, where an increasing Zeeman splitting isolates the electron spin from hyperfine-mediated pathways[33]. When $B_{\parallel}$ reaches 100 mT—well exceeding this characteristic threshold—the dephasing dip vanishes entirely into a single-exponential decay with a spin–lattice relaxation time of $T_1 = 0.77$ ns (Fig. 5d, red curve), aligning with the fundamental limits imposed by local vibrational modes and dynamic fluctuations in coordination environments[34,35].

As discussed above, histidine does not displace the strongly chelated EDTA ligand to form a new inner-sphere $[\mathrm{Cu(His)}_x]$ complex, but rather engages in outer-sphere supramolecular interaction or weak axial perturbation. Introducing 10 mM His conspicuously shifts the zero-field dephasing dip toward longer delay times ($\sim$ 1.83 ns;

Fig. 5b and inset). This delayed minimum reveals an increased $T_\Delta$ driven by non-covalent modulation of the local hyperfine environment, demonstrating a probe sensitivity to surrounding nuclear environments akin to bioinorganic spin sensing[36]. Concomitantly, at $100$ mT, the longitudinal spin lifetime increases from $0.77$ ns to $0.90$ ns (Fig. 5d). As established in copper complexes and metalloproteins, electron spin–lattice relaxation is dictated by resonant coupling to local vibrational modes and rotational dynamics[37]. Histidine coordination to the axial/outer coordination sphere stiffens the immediate supramolecular environment and increases the local rotational correlation time, thereby damping low-frequency vibrational motions and modulating dynamic spin–lattice pathways[35,37,38]. These results demonstrate that room-temperature coherence in solution is not an insurmountable casualty of bulk protic solvents, but a chemically tunable property governed by the local nuclear shell.

## Conclusion

In summary, we have demonstrated that rational ligand design can transform a magneto-optically sluggish aqua ion into an optically addressable, room-temperature molecular spin probe capable of all-optical coherence. By lowering the local coordination symmetry and lifting Laporte restrictions on d–d transitions, EDTA chelation markedly enhances optical transition dipole moments, enabling all-optical coherent spin readout of aqueous Cu(II) down to the micromolar regime. Magnetic-field-dependent spin dynamics further unravel the dominant dephasing mechanisms: transverse fields induce a crossover from inner-sphere nuclear hyperfine dephasing at low fields to Zeeman-amplified $g$-factor inhomogeneous broadening at elevated fields, whereas longitudinal fields progressively decouple the electron spin from the local nuclear bath. Together with the distinct coherence modulation observed for histidine over glycine, these findings establish that all-optical solution-state spin coherence can chemoselectively discriminate diamagnetic analytes, serving as a responsive reporter of local coordination events. While translating this platform to complex biological media and establishing a comprehensive quantitative structure-activity relationship across broader amino-acid libraries warrant further investigation, this work provides a vital proof-of-principle for chemoselective coordination-responsive spin control via an all-optical strategy in 3d transition-metal complexes and paves the way for microwave-free molecular quantum sensing in complex environments.

## Methods

### Materials and sample preparation

Anhydrous copper(II) ethylenediaminetetraacetate disodium salt $Na_2[Cu(EDTA)]$ (Cu–EDTA, ≥98%), anhydrous copper(II) sulfate ($CuSO_4$, ≥99%), $D_2O$ (≥99.9%), L-histidine (≥99.5%), and glycine (≥99%) were purchased from Aladdin and used as received without further purification. All samples were prepared in ultrapure water (or $D_2O$ where specified) without degassing.

The working concentrations of pristine copper solutions were chosen based on optical density and signal-to-noise requirements: 0.02–2 mM Cu–EDTA for spin dynamics, 2 and 50 mM Cu–EDTA and 50 mM $CuSO_4$ for steady-state absorption, and 50 mM (Cu–EDTA and $CuSO_4$) for MCD spectroscopy. For amino acid coordination studies, samples were prepared directly in the optical cuvettes by mixing aqueous Cu–EDTA with variable concentrations of L-histidine or glycine. Across all mixtures, the final concentration of Cu–EDTA was fixed at 2 mM, while the final amino acid concentrations spanned 0–100 mM. Specific concentrations for each experiment are indicated in the corresponding figure captions.

**Polarisation-sensitive pump–probe spectroscopy**

Electron spin dynamics were probed using polarisation-sensitive pump–probe spectroscopy in a non-degenerate transmission geometry (Fig. 2a). The laser source was an integrated Yb:KGW femtosecond laser amplifier (Pharos, Light Conversion) operating at a central wavelength of 1026 nm and a repetition rate of 50 kHz. The amplifier output was divided into two optical paths. One portion pumped a femtosecond optical parametric amplifier (OPA) to generate 750 nm pump pulses (spectral bandwidth ~117 $cm^{-1}$, pulse duration ~180 fs), which were then directed through a linear polariser and an electro-optic modulator (EOM) coupled with a quarter-wave plate ($\lambda/4$) to switch the circular polarisation between $\sigma^+$ and $\sigma^-$ at 850 Hz. The other portion passed through a second-harmonic bandwidth compressor and a picosecond OPA to yield 690 nm probe pulses (spectral bandwidth ~12.6 $cm^{-1}$, pulse duration ~3 ps). The relative time delay between the pump and probe pulses was adjusted using a motorised mechanical delay line.

Both beams were focused to spot diameters of ~300 μm at the sample position, verified by knife-edge measurements. The average pump power was fixed at 200 mW (pulse energy of 4 μJ), corresponding to the maximum available output from the OPA at 750 nm. For Cu–EDTA solutions at concentrations of 0.1–2 mM, an average probe power of 20 mW (pulse energy of 0.4 μJ) was used, whereas for more dilute solutions (0.02 mM) with weaker optical absorption, the probe power was elevated to 100 mW (pulse energy of 2 μJ) to improve the signal-to-noise ratio.

To isolate the subtle transient polarisation changes induced by spin dynamics without saturating the photodetectors, a crossed-polariser extinction geometry was implemented. Before the sample, a portion of the linearly polarised probe beam was split by a beam splitter (BS) to serve as an optical reference. After traversing the sample cuvette, residual pump scattering was blocked by spectral filters, and the probe beam was directed through a Glan-type analyser set to the crossed orientation relative to the incident probe polarisation to minimise transmission in the unperturbed state. Transient spin polarisation induced by the pump pulse alters the probe polarisation state, leading to a transmission change through the analyser. The transmitted signal and the reference beam were detected by a balanced photoreceiver, with a half-wave plate ($\lambda/2$) and polariser in the reference path to match intensity levels and eliminate common-mode laser intensity noise. The differential output was demodulated by a lock-in amplifier referenced to the 850 Hz EOM modulation with an integration time constant of 1 s.

To systematically eliminate non-spin contributions (such as thermal lensing, residual isotropic transient absorption, and linear birefringence or dichroism artifacts) and further enhance the signal-to-noise ratio, measurements were sequentially acquired with the fast axis of the pump $\lambda/4$ waveplate set to $+45°$ and $-45°$ relative to the incident linear polarisation. Because reversing the waveplate angle inverts the handedness of the modulated circular polarisations, the pure spin-dependent signal was extracted by subtracting the $-45°$ trace from the $+45°$ trace ($S_{\mathrm{final}} = S_{+45°} - S_{-45°}$), thereby canceling any helicity-independent background signals and instrumental artifacts. To further account for any residual non-spin contributions, a unified fitting model incorporating an oscillating spin coherence term, a non-oscillating background exponential decay term, and a constant offset was applied across all transient traces under transverse magnetic fields, while the main text highlights the core oscillatory envelope function for physical clarity.

All measurements were conducted at room temperature on molecular solutions contained in 1 mm path-length quartz cuvettes. External magnetic fields were supplied by an electromagnet. Transverse magnetic fields ($B_{\perp}$) up to 1,000 mT were applied perpendicular to both the light propagation axis and the pump polarisation vector. Longitudinal magnetic fields ($B_{\parallel}$) up to 100 mT were applied parallel to the light propagation direction.

**Steady-state UV–vis and MCD spectroscopy**

MCD and accompanying steady-state absorption spectra were recorded simultaneously at room temperature using a circular dichroism spectrometer (JASCO J-1500) equipped with a permanent magnet providing an external magnetic field of 1.6 T.

**Computational methods**

Density functional theory (DFT) and time-dependent DFT (TD-DFT) calculations were performed at the PBE0[39]-D4[40]/ma-def2-TZVP(-f)[41] and TD-PBE0[39]/ma-def2-TZVP[41] levels, respectively, with SMD implicit water solvation[42] in ORCA 6.1.1[43]. MCD spectra were calculated at the multireference QD-SC-NEVPT2/ma-def2-TZVP/SMD(water) level[44–46] with spin–orbit coupling included within the quasi-degenerate perturbation theory framework. Laplacian bond order (LBO) analyses and wavefunction-based hole–electron analyses were performed using Multiwfn 3.8[47,48] to evaluate the covalent bonding characteristics in the ground state and the nature of electronic transitions, respectively. Simulated UV–vis absorption spectra were likewise generated using Multiwfn 3.8[47]. Detailed computational protocols, active space definitions, and spectral broadening parameters are provided in the Supplementary Information.

**Acknowledgements:** This work was supported by the National Natural Science Foundation of China (Grant Nos. U25D8003, 12174108, 92461301, and 12274128) and the Science and Technology Commission of Shanghai Municipality (Grant No. 25JC3202500). We thank the ECNU Multifunctional Platform for Innovation (001) and the HPC Research Computing Team for providing computational and storage resources.

**Author contributions:** J.L. and Q.W. carried out the experiments, collected the data, and performed the data analysis. Q. S., Z. H., and H. S. conducted quantum chemistry calculations. D.F. and Z.S. conceived and supervised the project. All authors discussed the results, contributed to their interpretation, and participated in writing and revising the manuscript.

**Extended Data figures**

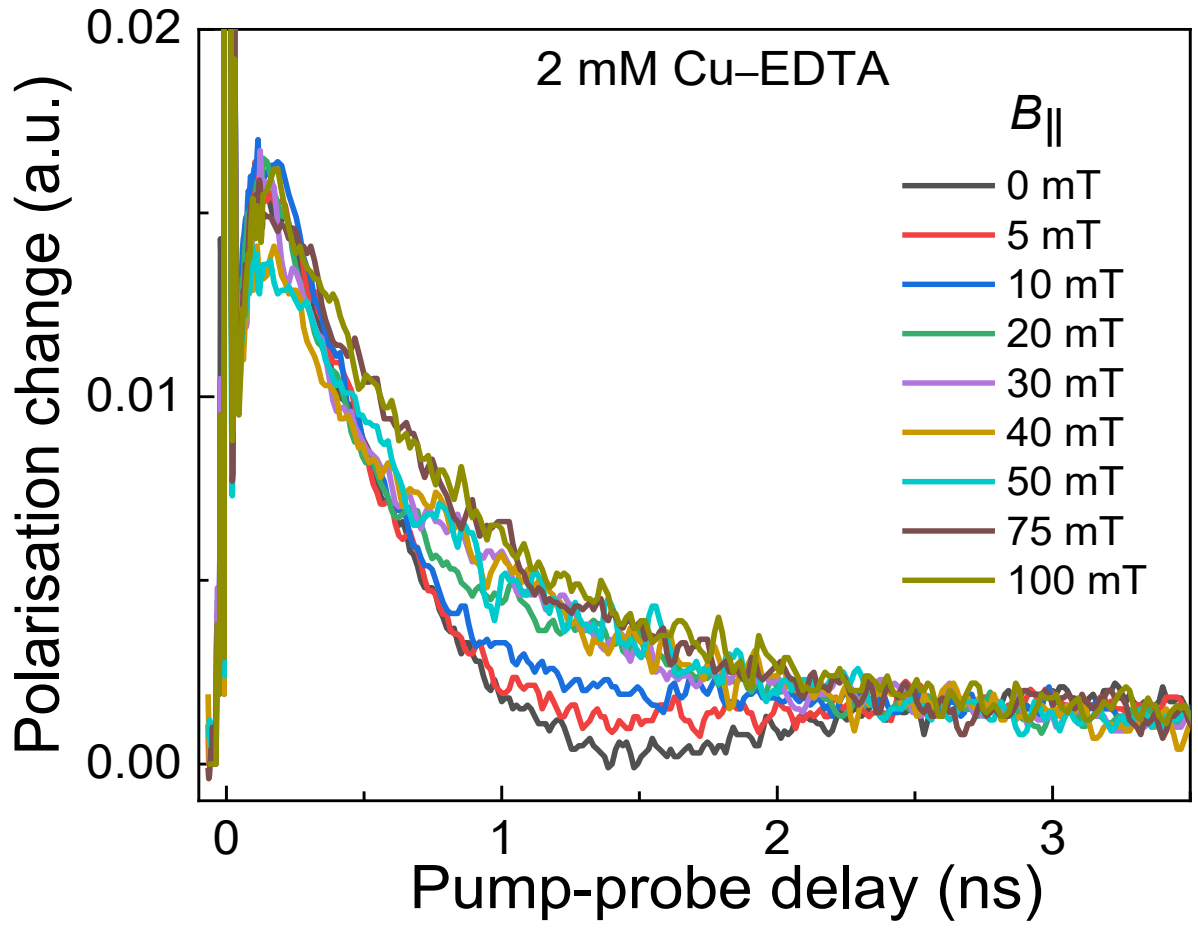

**Extended Data Fig. 1 | Magnetic-field dependence of spin dynamics in aqueous Cu–EDTA.** Spin dynamics of 2 mM Cu–EDTA measured under longitudinal magnetic fields ($B_{\parallel}$) from 0 to 100 mT. This full dataset serves as the basis for the averaged signal intensities presented in Fig. 5c.

## Data availability

The data that support the findings of this study are available within the Article and its Supplementary Information. Source data for the figures are available from the corresponding authors upon reasonable request.

## Competing interests

The authors declare no competing interests.

# Supplementary Information

## Supplementary Notes

### Supplementary Note 1. Computational details

The geometries of $[Cu(EDTA)]^{2-}$, $[Cu(EDTA)(H_2O)]^{2-}$, and $[Cu(H_2O)_5]^{2+}$ were optimized at the PBE0[1]-D4[2]/ma-def2-TZVP(-f)[3] level, with solvent effects described using the SMD implicit solvation model for water[4]. The vertical excitation energies and oscillator strengths for 20 states were calculated at the TD-PBE0/ma-def2-TZVP/SMD(water) level based on the optimized geometries. The simulated absorption spectra were generated using the Multiwfn 3.8 program[5] by applying Gaussian broadening with a full width at half maximum (FWHM) of 0.5 eV, and selected excited states were further characterized by real-space hole–electron analysis using the corresponding module implemented in Multiwfn[6]. To quantify the covalent bonding characteristics in the primary coordination sphere, Laplacian bond order (LBO) analyses were performed on the ground-state wavefunctions using Multiwfn 3.8[5], alongside coordination bond length evaluations on the optimized geometries.

Magnetic circular dichroism (MCD) spectra were calculated at the QD-SC-NEVPT2[7,8]/ma-def2-TZVP/SMD(water) level. The multireference wave functions were described using a CAS(9,5) active space comprising nine electrons in five orbitals, with the CASSCF[9] orbitals state-averaged over the lowest five doublet roots. Dynamic electron correlation was treated using the strongly contracted N-electron valence state second-order perturbation theory (SC-NEVPT2), followed by a quasi-degenerate treatment using the Hermitian Van Vleck effective Hamiltonian. Spin–orbit coupling (SOC) was subsequently included within the quasi-degenerate perturbation theory framework. The MCD calculations were performed at 298.15 K under an external magnetic field of 1.6 T. Electric-dipole transition moments were evaluated in both the length and velocity representations, and the dipole-length results were used for spectral analysis. Theoretical MCD band profiles were simulated with a Gaussian FWHM of 0.4 eV. All electronic-structure calculations were performed using the ORCA 6.1.1 package[10].

**Supplementary Note 2. Coordination models and dynamic equilibrium in aqueous solution**

As revealed by X-ray absorption spectroscopy[11,12], the square-pyramidal $[Cu(H_2O)_5]^{2+}$ complex represents the time-averaged, energetically dominant species in aqueous solution and was therefore adopted as the representative model for unchelated Cu(II) ions. For the Cu(II)–EDTA complex in pure aqueous solution, a single static structure cannot account for the observed solution-state properties. Proton nuclear magnetic relaxation dispersion studies by Oakes and Smith established that aqueous Cu(II)–EDTA exists in a dynamic coordination equilibrium between a fully chelated hexadentate form ($[Cu(EDTA)]^{2-}$) and an aquo-pentadentate form ($[Cu(EDTA)(H_2O)]^{2-}$)[13]:

$$[Cu(EDTA)]^{2-} + H_2O \rightleftharpoons [Cu(EDTA)(H_2O)]^{2-}$$

In the aquo-pentadentate conformer, one carboxylate arm is detached from the copper center and an apical position is occupied by a coordinated water molecule. Across the broad neutral-to-basic pH plateau (pH 4–10), the average inner-sphere water coordination number was determined to be $n \approx 0.38$, yielding a thermodynamic population ratio of 62% hexadentate $[Cu(EDTA)]^{2-}$ and 38% aquo-pentadentate $[Cu(EDTA)(H_2O)]^{2-}$ at room temperature[13]. We adopt this 62:38 distribution as a representative literature benchmark to capture the dynamic hydration ensemble. While minor variations in the actual equilibrium fraction may arise in unbuffered aqueous solutions, our conclusions remain unaffected: as shown in Supplementary Table 1, both pure conformers independently exhibit marked absorption and MCD enhancements over $[Cu(H_2O)_5]^{2+}$.

**Supplementary Note 3. Hole–electron analysis and metal–ligand covalency**

To uncover the electronic origin of the enhanced transition probabilities and dipole moments, real-space hole–electron analyses were performed using Multiwfn[5,6]. For both the unchelated aqua ion and the chelated Cu(II)–EDTA species, the analysis focused on the three ligand-field (d–d) transitions that dominate the visible absorption envelope (denoted as States 1–3 in Supplementary Fig. 1a–c, corresponding to transitions from the ground state to the second through fourth excited states; the lowest-energy first excited state, predicted theoretically above 1150 nm, lies well beyond the experimental visible/NIR detection window). In the ligand-field hole formalism of $d^9$ Cu(II), these transitions describe the excitation of the hole from the ground $d_{x^2-y^2}$ state into the higher-lying $d_{xy}$, $d_{xz}$, and $d_{yz}$ hole states.

In $[Cu(H_2O)_5]^{2+}$, both the hole and the excited electron remain predominantly localized within the central copper 3d manifold (Supplementary Fig. 1c), reflecting an essentially localized, ionic d–d transition. Concomitant with the pseudo-centrosymmetric ligand field in aqueous solution, this spatial confinement reinforces the Laporte forbiddenness, leading to vanishingly small transition electric dipole moments and weak oscillator strengths. In sharp contrast, for both $[Cu(EDTA)]^{2-}$ and $[Cu(EDTA)(H_2O)]^{2-}$, the hole distributions exhibit substantial real-space delocalization across the chelating carboxylate oxygen donor atoms (Supplementary Fig. 1a,b). This increased ligand participation in the electronic excitations signifies enhanced metal–ligand orbital mixing.

To further evaluate the bonding nature underlying this electronic structure, Laplacian bond order (LBO) analysis was conducted on the ground-state coordination spheres (Supplementary Fig. 1d–f). For $[Cu(H_2O)_5]^{2+}$, the equatorial Cu–O coordination bonds exhibit LBO values in the range of 0.158–0.166 with bond lengths of ~1.98–1.99 Å (and an axial LBO of 0.069 at 2.219 Å). In comparison, the primary equatorial carboxylate Cu–O bonds in $[Cu(EDTA)]^{2-}$are appreciably contracted to 1.950 Å and display enhanced LBO values of ~0.190, alongside covalent contributions from the coordinated amine nitrogens (LBO $\approx$ 0.132). In addition, for $[Cu(EDTA)(H_2O)]^{2-}$, the equatorial carboxylate Cu–O bonds also remain significantly contracted (1.94–1.95 Å) with high LBO values of 0.189–0.198, while the two Cu–N bonds show pronounced asymmetry with LBO values of 0.236 (2.088 Å) and 0.066 (2.321 Å), respectively. The higher overall coordination bond orders corroborate stronger covalent bonding interactions upon EDTA chelation.

In synergy with the pronounced coordination asymmetry imposed by the multidentate chelate framework, this augmented metal–ligand orbital mixing relaxes the Laporte restriction and mixes metal 3d states with charge-transfer character, substantially enlarging the transition dipole moments (with computed $D^2$ values reaching 0.0212 au$^2$ for $[Cu(EDTA)]^{2-}$). These electronic and bonding factors collectively explain how coordination engineering by EDTA yields a ~7-fold amplification in molar absorptivity and a ~6-fold enhancement in peak MCD amplitude despite retaining a nominal $d^9$ Cu(II) configuration.

**Supplementary Note 4. Comparison between theoretical predictions and experimental data**

A summary of calculated optical and magneto-optical properties (relative to calculated $[Cu(H_2O)_5]^{2+}$ alongside experimental enhancement ratios (relative to aqueous $CuSO_4$) is presented in Supplementary Table 1:

- Model 1 (100% $[Cu(EDTA)]^{2-}$): Predicts a 14.6-fold enhancement in absorption and a 4.4-fold enhancement in MCD intensity.
- Model 2 (100% $[Cu(EDTA)(H_2O)]^{2-}$): Predicts a 7.3-fold enhancement in absorption and a 2.1-fold enhancement in MCD intensity.
- Model 3 (Weighted ensemble: 62% Model 1 + 38% Model 2): Predicts an effective absorption enhancement of 11.2-fold and an MCD enhancement of 3.5-fold.

We note that the computed vertical excitation energies of the ligand-field transitions for $[Cu(H_2O)_5]^{2+}$ are appreciably blueshifted relative to the broad experimental absorption band maximum (~800 nm). Such discrepancies are well known in static cluster TD-DFT calculations of open-shell transition-metal aqua complexes, primarily originating from the omission of explicit second-sphere hydrogen-bonding networks and dynamic Jahn–Teller conformational fluctuations in aqueous solution. Because the primary objective of this work is to resolve the relative spectral amplification induced by EDTA chelation rather than absolute transition energies, this systematic offset does not compromise the underlying electronic mechanism.

Furthermore, for $[Cu(H_2O)_5]^{2+}$, the d–d transitions are Laporte-forbidden and experimentally borrow intensity via asymmetric vibronic coupling. Because static TD-DFT does not explicitly account for vibronic borrowing at the non-vibrating equilibrium geometry, it systematically underestimates the baseline oscillator strength of the pseudo-centrosymmetric aqua complex, thereby overestimating the relative absorption enhancement ratio for Model 3 (11.2× calculated vs. 7.3× experimental). Nonetheless, the weighted thermodynamic ensemble captures the overall magnitude and trend of the experimental benchmarks (7.3× in absorption and 5.8× in MCD), validating that the pronounced coordination asymmetry and enhanced metal–ligand orbital mixing govern the observed spectral amplification.

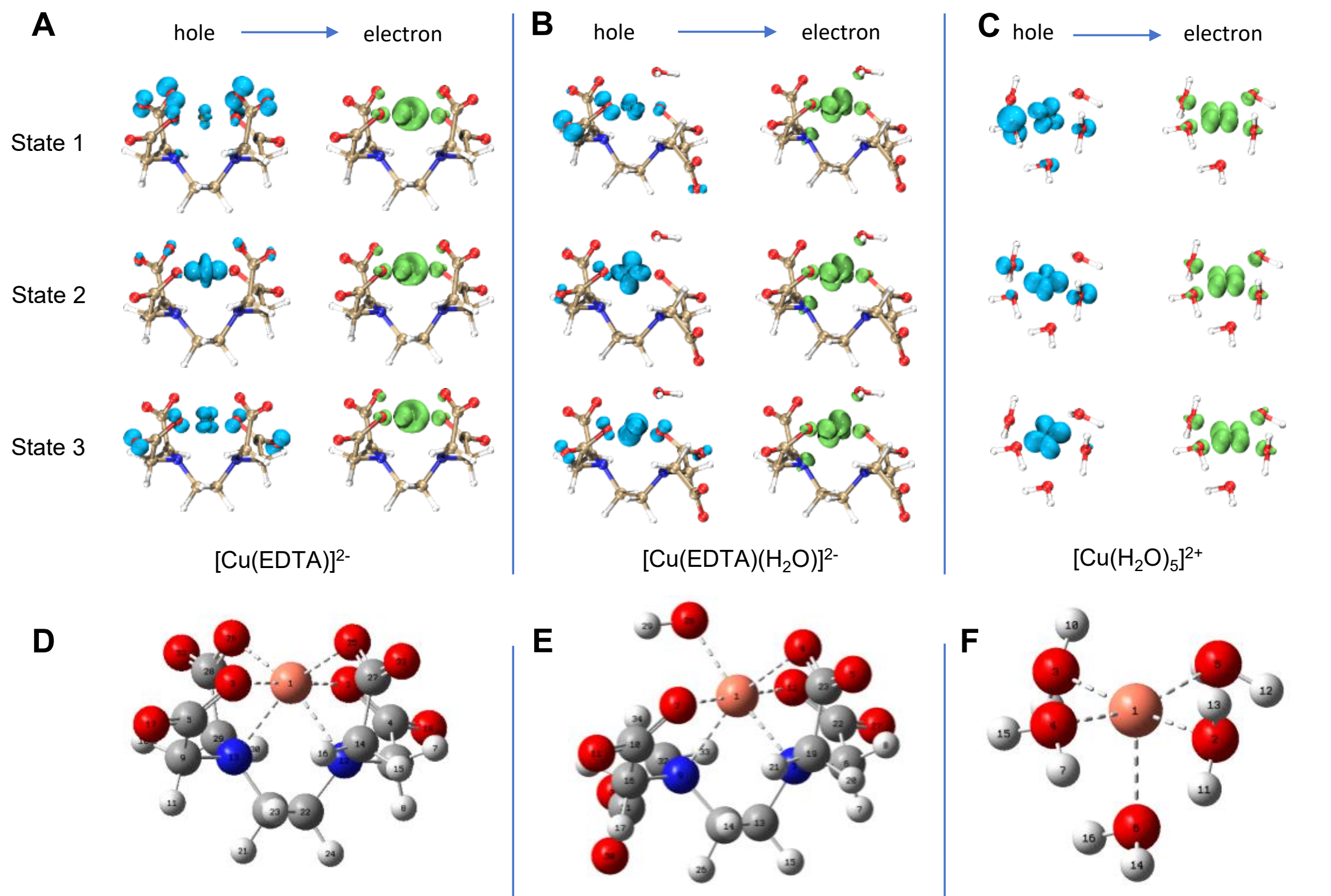

| | Bond(Å) | LBO |
|---|---|---|
| O2–Cu1 | 1.94974 | 0.189597 |
| O3–Cu1 | 1.94973 | 0.189606 |
| N12–Cu1 | 2.23701 | 0.132408 |
| N13–Cu1 | 2.23715 | 0.132298 |
| O25–Cu1 | 2.13798 | 0.097964 |
| O26–Cu1 | 2.13775 | 0.098053 |

| | Bond (Å) | LBO |
|---|---|---|
| O2–Cu1 | 1.95220 | 0.188773 |
| N3–Cu1 | 2.08824 | 0.235792 |
| O4–Cu1 | 2.32498 | 0.049160 |
| N9–Cu1 | 2.32124 | 0.066451 |
| O12–Cu1 | 1.94299 | 0.197508 |
| O28–Cu1 | 2.08531 | 0.115374 |

| | Bond(Å) | LBO |
|---|---|---|
| O2–Cu1 | 1.99465 | 0.158474 |
| O3–Cu1 | 1.99343 | 0.158872 |
| O4–Cu1 | 1.98868 | 0.164239 |
| O5–Cu1 | 1.98353 | 0.166282 |
| O6–Cu1 | 2.21907 | 0.069056 |

**Supplementary Fig. 1 | Hole–electron analysis and metal–ligand covalency of copper complexes. a–c,** Real-space distributions of the hole (blue) and electron (green) for visible ligand-field transitions to the second through fourth excited states (denoted as States 1–3) in [Cu(EDTA)]$^{2-}$ (**a**), [Cu(EDTA)($H_2O$)]$^{2-}$ (**b**), and [Cu($H_2O$)$_5$]$^{2+}$ (**c**). Blue arrows indicate the transition direction from hole to electron. All isosurfaces are plotted at a uniform isovalue of 0.01 a.u. **d–f**, Optimized coordination geometries alongside calculated coordination bond lengths (in Å) and Laplacian bond orders (LBO) for [Cu(EDTA)]$^{2-}$ (**d**), [Cu(EDTA)($H_2O$)]$^{2-}$ (**e**), and [Cu($H_2O$)$_5$]$^{2+}$ (**f**).

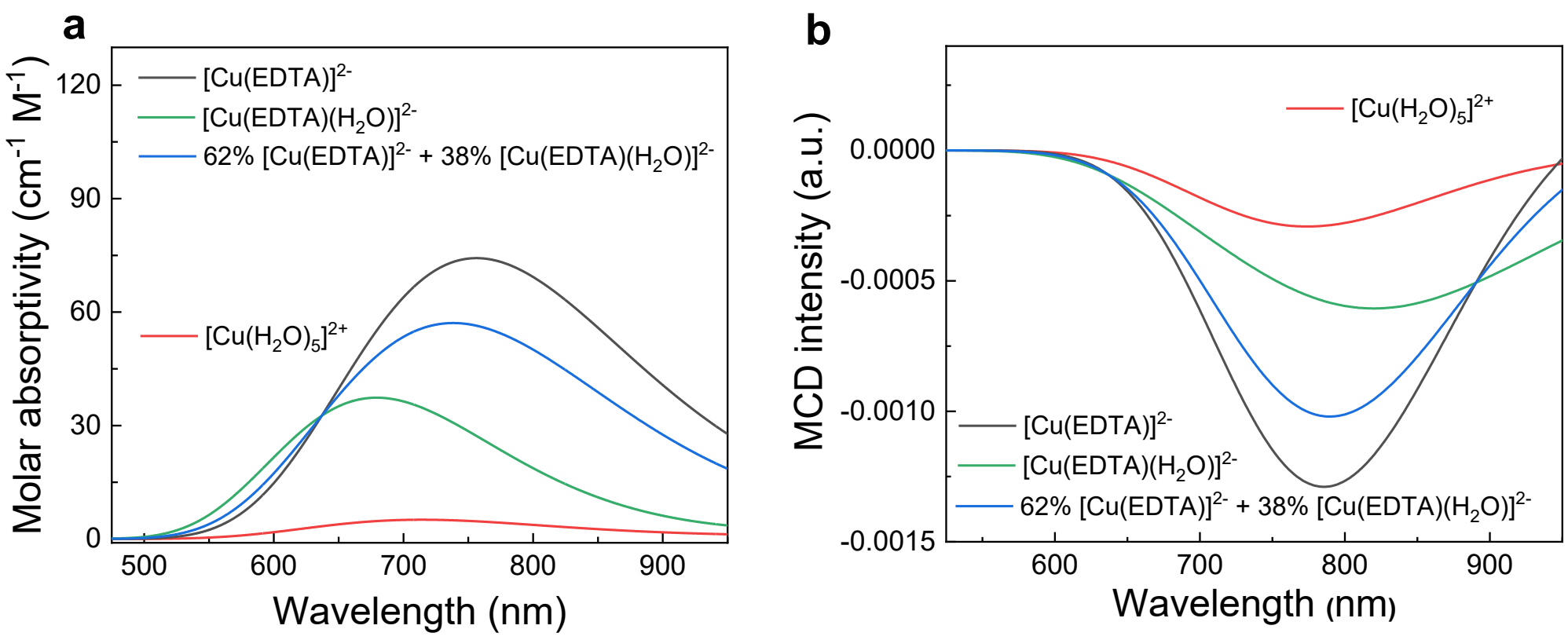

**Supplementary Fig. 2 | Simulated UV–vis absorption and magnetic circular dichroism (MCD) spectra of copper species. a,b**, Theoretical absorption spectra (**a**, Gaussian FWHM = 0.5 eV) and MCD spectra (**b**, Gaussian FWHM = 0.4 eV) calculated for the fully chelated hexadentate complex $[Cu(EDTA)]^{2-}$ (black), the aquo-pentadentate complex $[Cu(EDTA)(H_2O)]^{2-}$ (green), their thermodynamic equilibrium mixture (62% $[Cu(EDTA)]^{2-}$ + 38% $[Cu(EDTA)(H_2O)]^{2-}$, blue), and the unchelated aqua ion $[Cu(H_2O)_5]^{2+}$ (red). Absorption profiles were calculated at the TD-PBE0/ma-def2-TZVP/SMD(water) level, and MCD spectra were computed at the QD-SC-NEVPT2/ma-def2-TZVP/SMD(water) level under an external magnetic field of 1.6 T at 298.15 K.

**Supplementary Table 1 |** Comparison of experimental and calculated absorption and MCD enhancement ratios of Cu–EDTA relative to aqueous $CuSO_4$.

| Sample/Model | Structural Description | Absorption Ratio ($Cu–EDTA/CuSO_4$) | MCD Ratio ($Cu–EDTA/CuSO_4$) |
|---|---|---|---|
| Experimental | Aqueous solution (RT) | 7.3 | 5.8 |
| Model 1 | 100% $[Cu(EDTA)]^{2-}$ | 14.6 | 4.4 |
| Model 2 | 100% $[Cu(EDTA)(H_2O)]^{2-}$ | 7.3 | 2.1 |
| Model 3 | 62% Model 1 + 38% Model 2 | 11.2 | 3.5 |

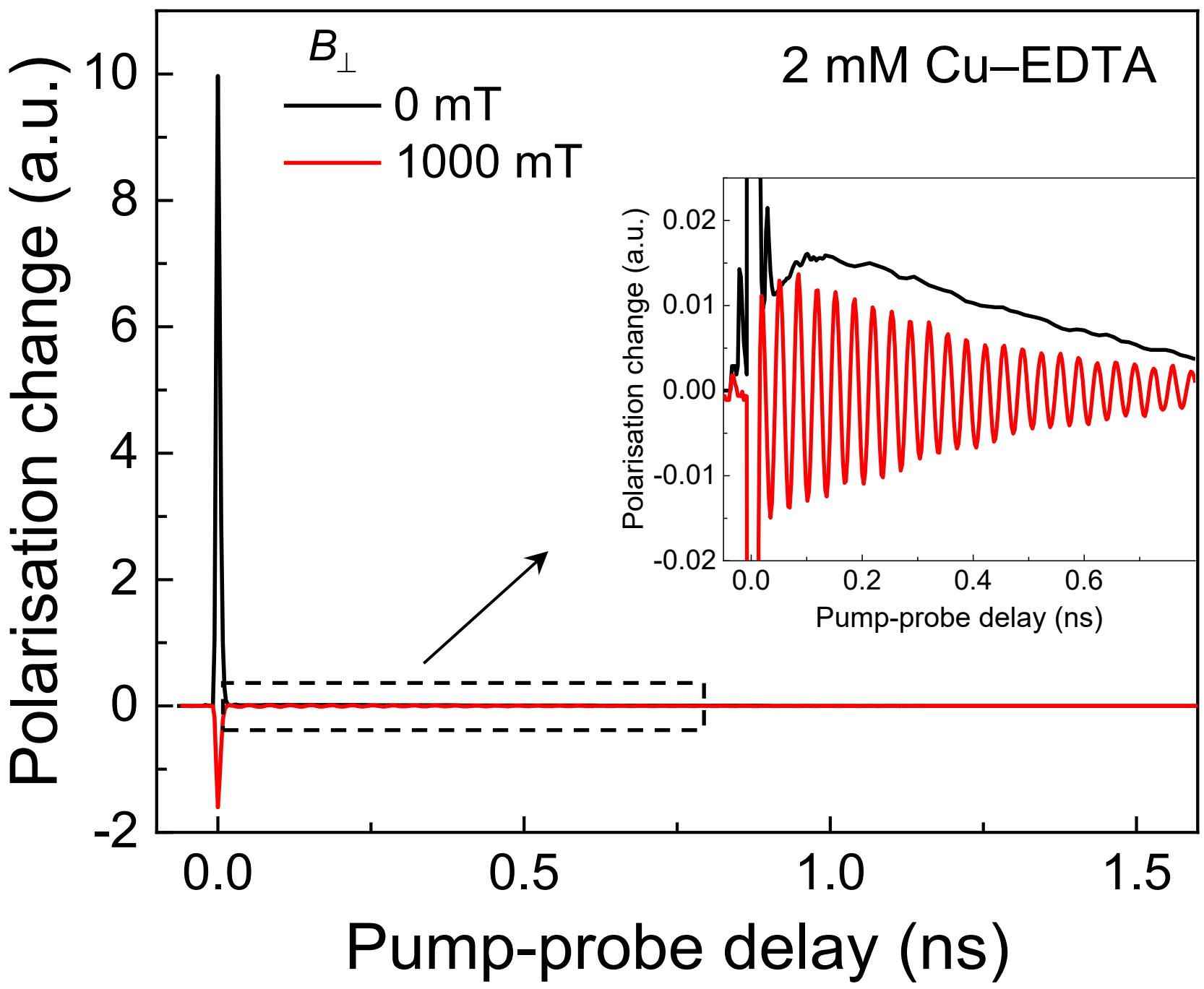

**Supplementary Fig. 3 | Time-resolved traces near zero time delay.** Polarization change traces of 2 mM Cu–EDTA measured at 0 mT and 1000 mT. The strong peak around zero delay originates from the instantaneous optical Kerr effect, which perturbs the initial spin coherence signal (as highlighted in the inset) and is excluded from subsequent kinetic fitting.

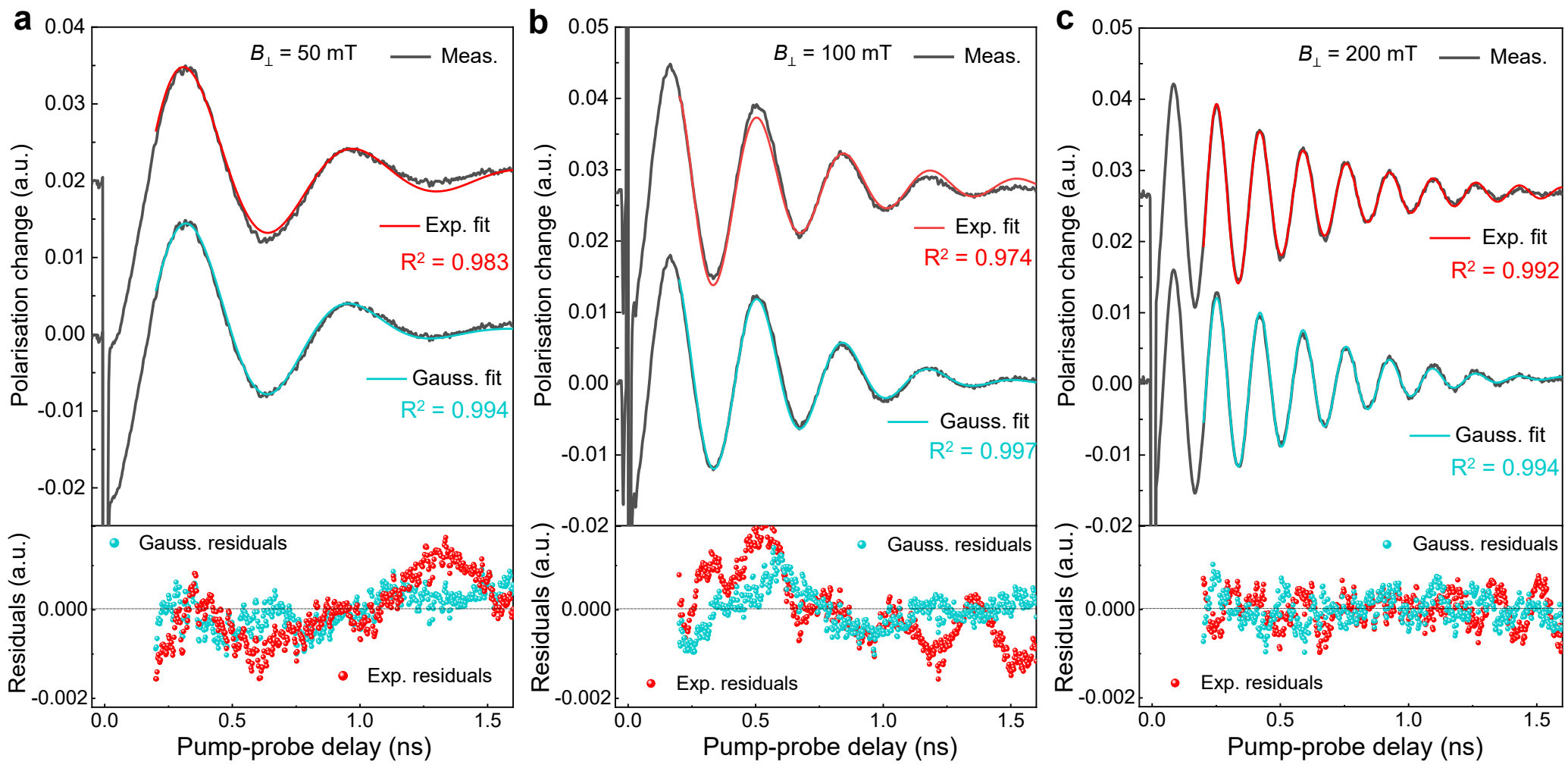

**Supplementary Fig. 4 | Comparison of fitting models for spin dephasing dynamics at low transverse magnetic fields. a–c**, Time-resolved polarisation changes measured under $B_\perp = 50$ mT (**a**), $100$ mT (**b**), and $200$ mT (**c**) (grey curves, vertically offset for clarity). Experimental data are fitted with a mono-exponential decay model $S(t) = S_0 \cos(2\pi\nu_L t) \exp(-t/T_2^*)$ (red curves) and a Gaussian decay model $S(t) = S_0 \cos(2\pi\nu_L t) \exp\left(-t^2/2{T_2^*}^2\right)$ (cyan curves). Lower panels show the corresponding residuals for the exponential (red) and Gaussian (cyan) fits. The Gaussian envelope provides superior fits—especially at lower fields (50 and 100 mT)—yielding reduced residual deviations and higher $R^2$ values across all fields.

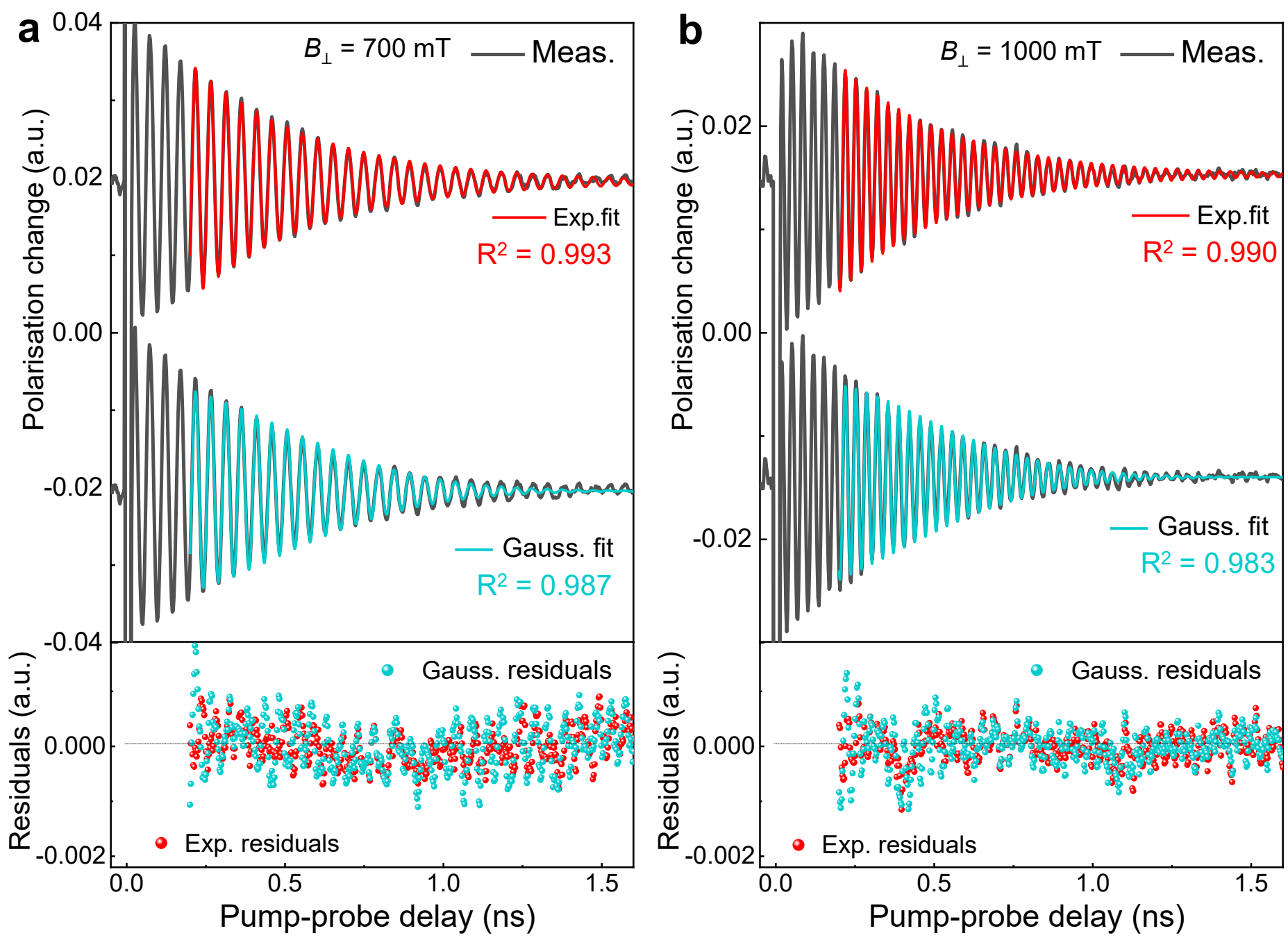

**Supplementary Fig. 5 | Comparison of fitting models for spin dephasing dynamics at high magnetic fields. a,b,** Time-resolved polarisation dynamics measured under $B_\perp = 700$ mT (**a**) and $1000$ mT (**b**) (grey curves, vertically offset for clarity). Data are fitted using the mono-exponential (red curves) and Gaussian (cyan curves) decay models defined in Supplementary Fig. 4. Lower panels show the corresponding fitting residuals for the exponential (red) and Gaussian (cyan) fits. In this high-field regime, the mono-exponential decay provides a markedly superior fit, yielding higher $R^2$ values and reduced residual deviations across both fields.

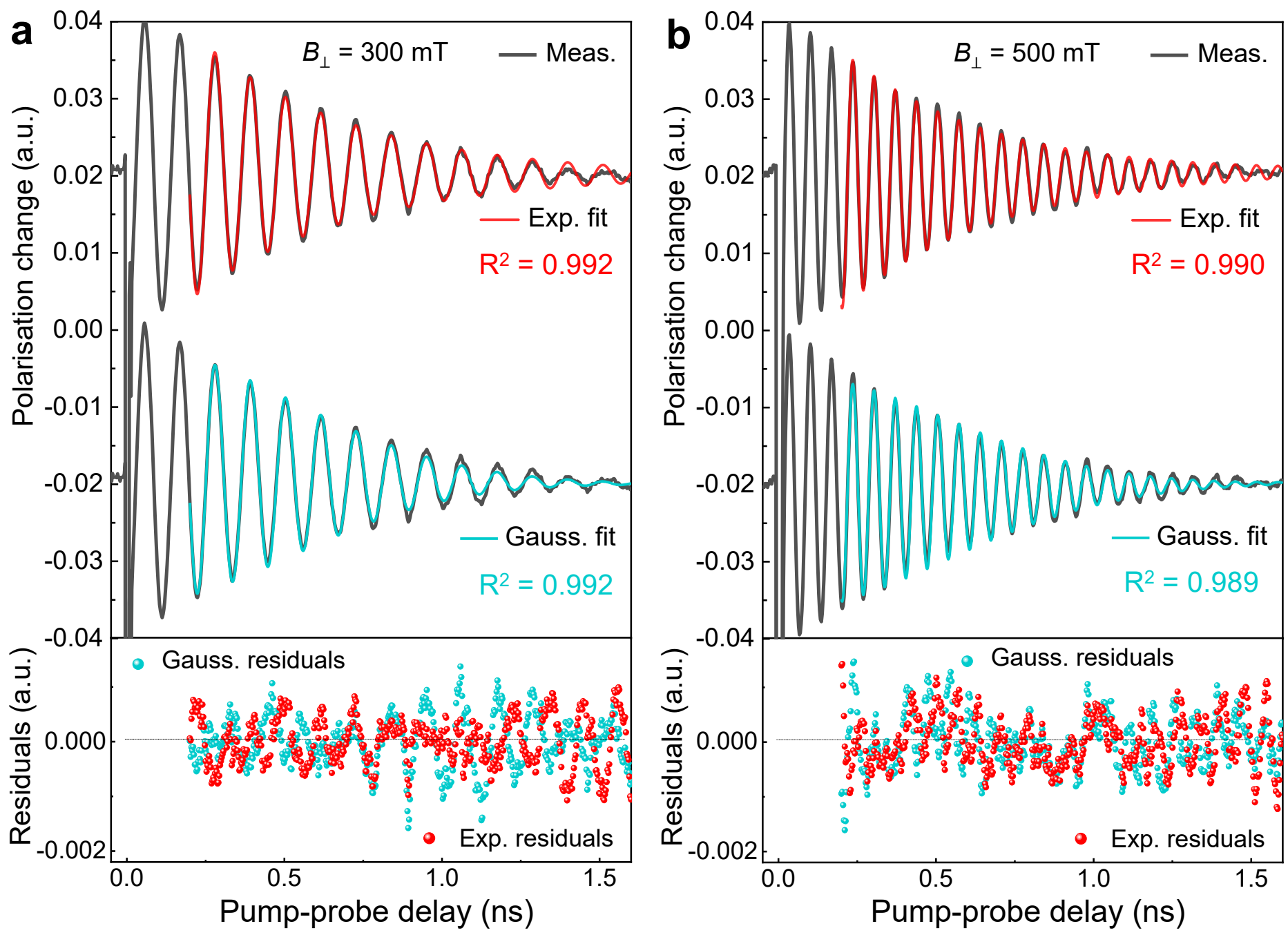

**Supplementary Fig. 6 | Comparison of fitting models for spin dephasing dynamics at intermediate magnetic fields. a**,**b**, Time-resolved polarisation dynamics measured under $B_\perp = 300$ mT (**a**) and 500 mT (**b**) (grey curves, vertically offset for clarity). Data are fitted using the mono-exponential (red curves) and Gaussian (cyan curves) decay models defined in Supplementary Fig. 4. Lower panels show the corresponding fitting residuals for the exponential (red) and Gaussian (cyan) fits. In this intermediate-field regime, both models exhibit nearly indistinguishable goodness-of-fit, yielding comparable $R^2$ values and residual deviations.

.

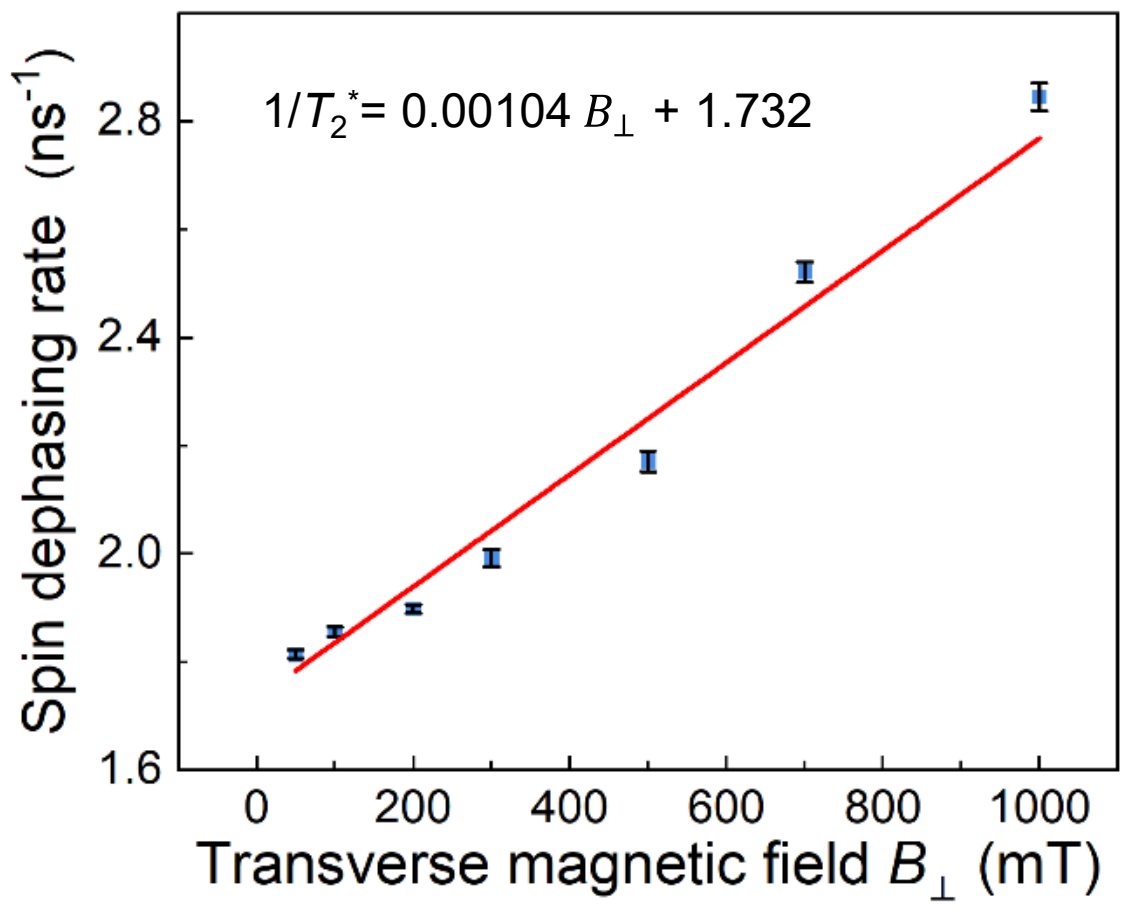

**Supplementary Fig. 7 | Magnetic-field dependence of spin dephasing rates.** Spin dephasing rate ($1/T_2^*$) of Cu–EDTA as a function of the transverse magnetic field ($B_\perp$). Blue squares represent experimental data points, and the solid red line shows the linear fit to the data ($1/T_2^* = 0.00104\,B_\perp + 1.732$). The intercept yields a zero-field spin dephasing time of $T_2^*(0) \approx 0.58$ ns ($1.732\ \text{ns}^{-1}$).

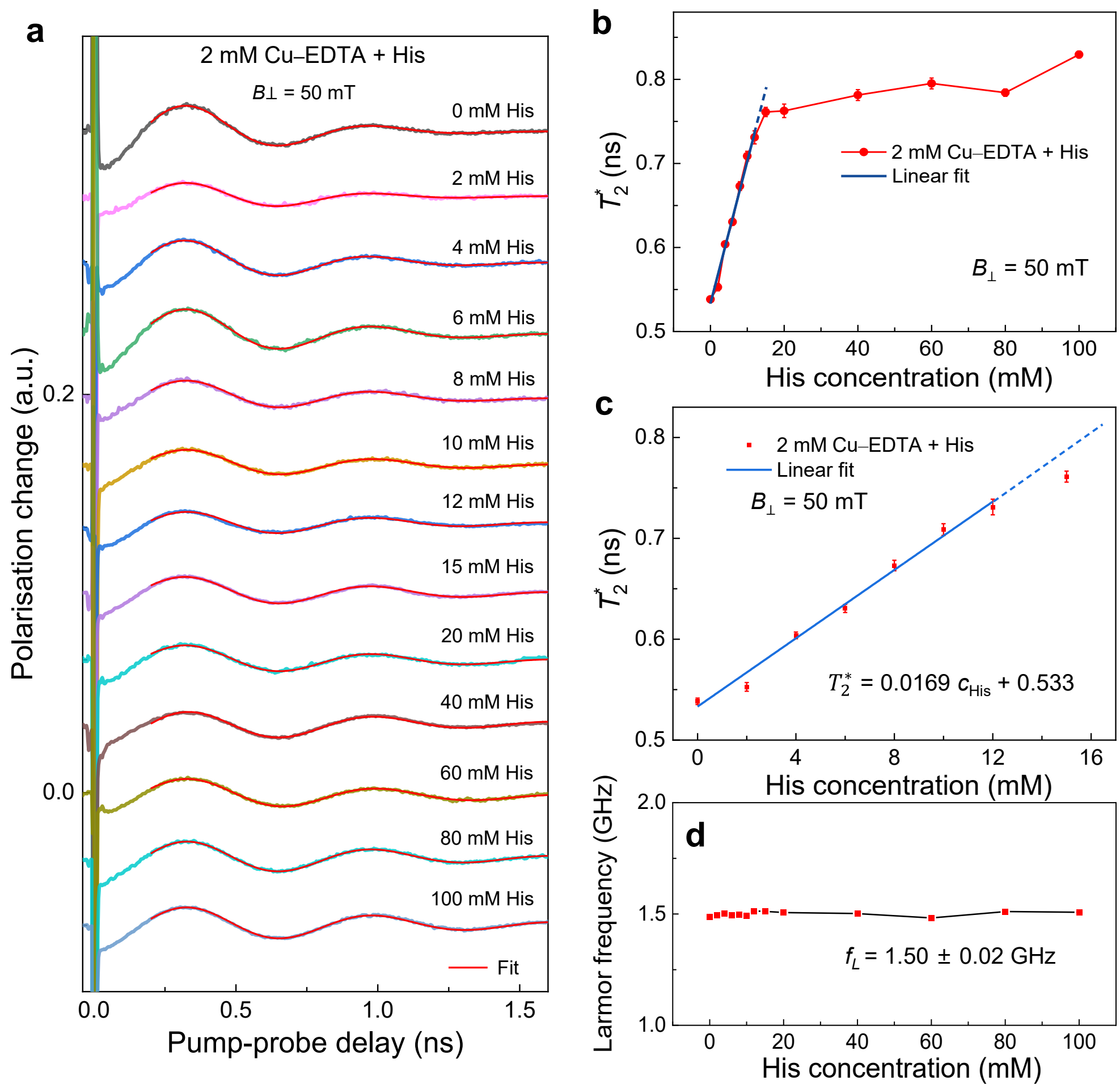

**Supplementary Fig. 8 | Spin dynamics and histidine-dependent dephasing in Cu–EDTA complexes. a,** Time-resolved polarisation change traces (colored lines) and corresponding fits (red lines) for 2 mM Cu–EDTA in the presence of varying histidine (His) concentrations ($B_{\perp} = 50$ mT). Traces are vertically offset for clarity. **b**, Extracted transverse spin dephasing time ($T_2^*$) as a function of His concentration. **c**, Zoom-in view of the low-concentration regime (0–15 mM) from **b**, showing the linear fit. **d**, Fitted Larmor precession frequencies ( $f_L$ ) across the tested His concentration range, remaining constant at $1.50 \pm 0.02$ GHz.

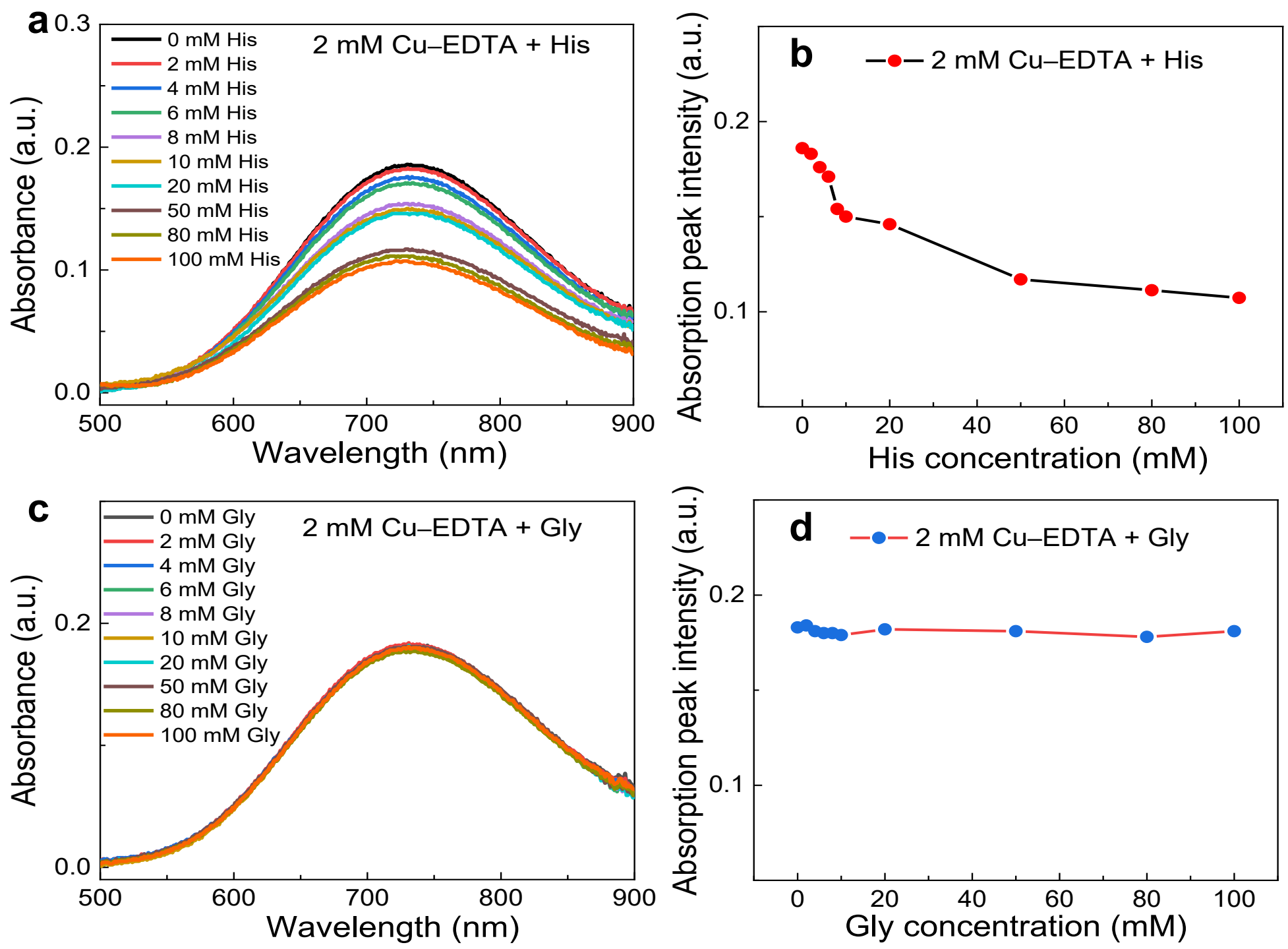

**Supplementary Fig. 9 | Concentration-dependent absorption spectra of Cu–EDTA with histidine and glycine. a,** UV–vis absorption spectra of Cu–EDTA (2 mM) in the presence of varying concentrations of histidine (His; 0–100 mM). **b**, Absorption peak intensity as a function of His concentration. **c**, UV–vis absorption spectra of Cu–EDTA (2 mM) in the presence of varying concentrations of glycine (Gly; 0–100 mM). **d**, Absorption peak intensity as a function of Gly concentration.

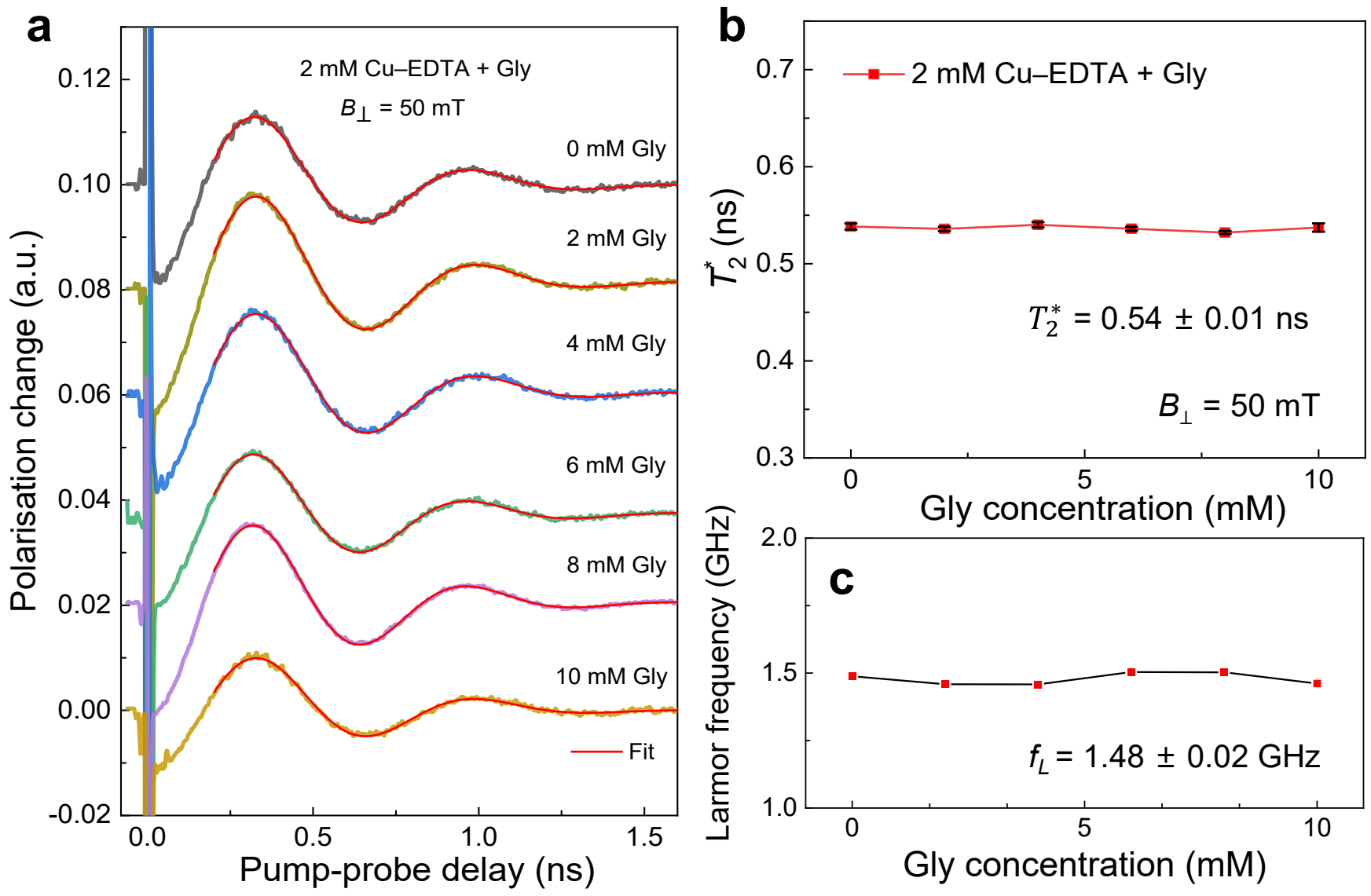

**Supplementary Fig. 10 | Spin dynamics and glycine-independent dephasing in aqueous Cu–EDTA. a,** Time-resolved polarisation change traces (colored lines) and corresponding fits (red lines) for 2 mM Cu–EDTA in the presence of varying glycine (Gly) concentrations ($B_{\perp} = 50$ mT). Traces are vertically offset for clarity. **b**, Extracted transverse spin dephasing time ($T_2^*$) as a function of Gly concentration, remaining essentially unchanged at $0.54 \pm 0.01$ ns. **c**, Fitted Larmor precession frequencies ($f_{\mathrm{L}}$) across the tested Gly concentration range, remaining constant at $1.48 \pm 0.02$ GHz.

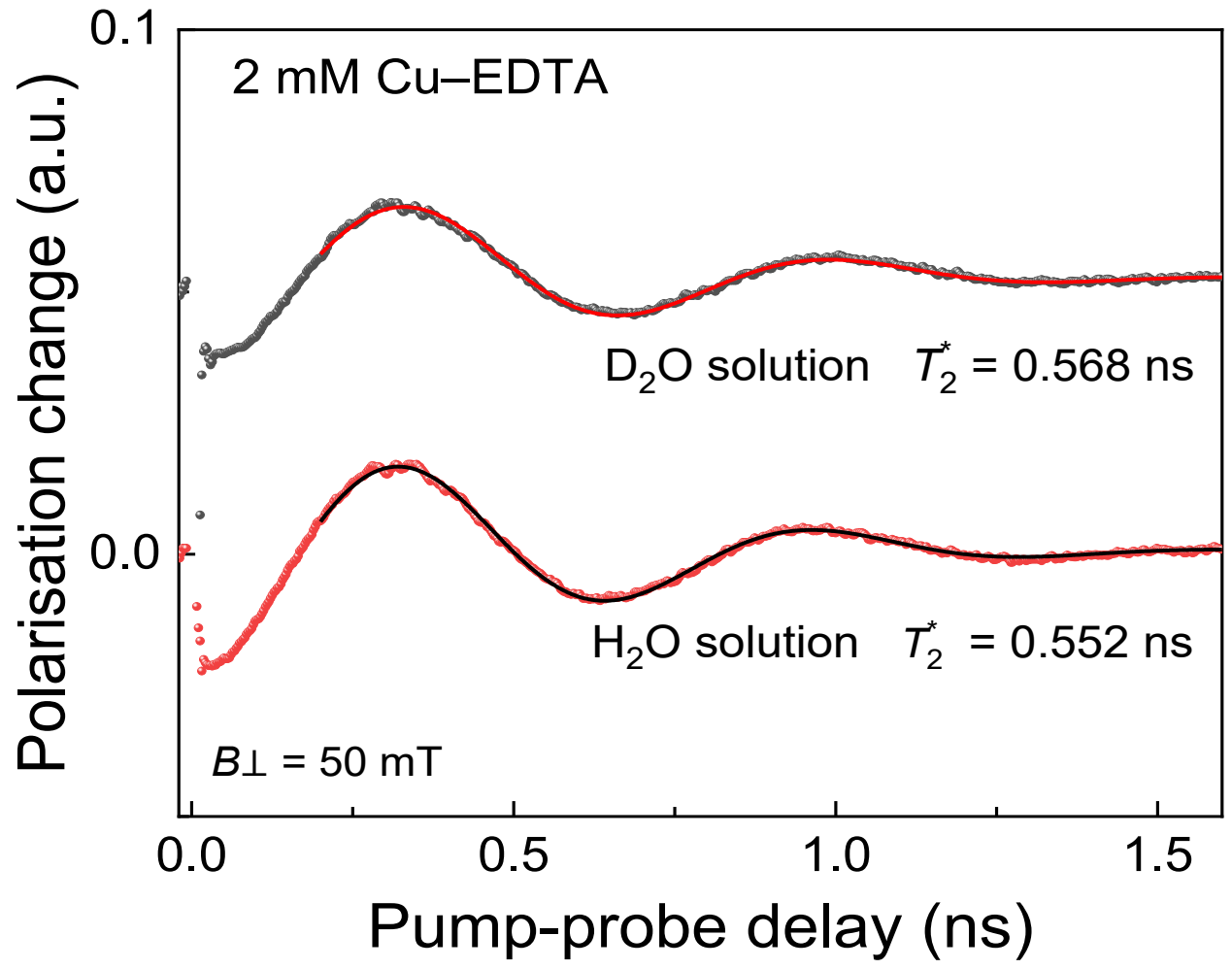

**Supplementary Fig. 11 | Spin coherence dynamics of Cu–EDTA in $H_2O$ and $D_2O$.** Time-resolved polarisation change measured for 2 mM Cu–EDTA in $D_2O$ (top) and $H_2O$ (bottom) under an external transverse magnetic field of $B_{\perp} = 50\ \mathrm{mT}$. Experimental data points are represented by dots (grey for $D_2O$, red for $H_2O$), and red/black solid curves denote fits to a Gaussian-damped cosine function, yielding effective dephasing times of $T_2^* = 0.568\ \mathrm{ns}$ and $0.552\ \mathrm{ns}$, respectively. Traces are vertically offset for clarity.

## Supplementary References